\documentclass[journal=jpcl,manuscript=letter]{achemso}
\usepackage[version=3]{mhchem}
\usepackage{caption}
\usepackage{subcaption}
\usepackage{hyperref}
\usepackage{natbib}
\usepackage{siunitx}
\usepackage{tabularx}
\usepackage{amsmath}
\usepackage{amssymb}
\usepackage{xcolor}
\usepackage{threeparttable}
\usepackage{multirow}
\usepackage{booktabs}
\usepackage{subcaption}
\usepackage{braket}
\usepackage{derivative}
\usepackage{physics}
\usepackage{gensymb}

\newcommand{\bfX}{{\mathbf{X}}}
\newcommand{\bfA}{{\mathbf{A}}}

\newcommand{\bfzero}{{\mathbf{0}}}

\newcommand{\qa}{\text{Q}}

\newcommand{\aobs}{\mathcal{O}}

\author{Arpan Kundu}
\affiliation{Pritzker School of Molecular Engineering, The University of Chicago, Chicago, Illinois 60637, United States}
\email{arpan.kundu@gmail.com}
\author{Giulia Galli}
\affiliation{Department of Chemistry, University of Chicago, Chicago, Illinois 60637, United States}
\alsoaffiliation{Pritzker School of Molecular Engineering, The University of Chicago, Chicago, Illinois 60637, United States}
\alsoaffiliation{Materials Science Division and Center for Molecular Engineering, Argonne National Laboratory, Lemont, Illinois 60439, United States}
\email{gagalli@uchicago.edu}

\title{Quantum Vibronic Effects on the Excitation Energies of the Nitrogen-Vacancy Center in Diamond}
\begin{document}
\maketitle

\let\thefootnote\relax\footnotetext{This document is the unedited Author’s version of a Submitted Work that was subsequently accepted for publication in {\it J. Phys. Chem. Lett}, copyright © 2024 American Chemical Society after peer review.\hfill}

\begin{abstract}
We investigate the impact of quantum vibronic coupling on the electronic properties of solid-state spin defects using  stochastic methods and first principles molecular dynamics with a quantum thermostat. Focusing on the negatively charged nitrogen-vacancy center in diamond as an exemplary case,  we found  a significant dynamic Jahn-Teller splitting of the doubly degenerate single-particle levels within the diamond's band gap, even at 0 K, with a magnitude exceeding 180 meV.  This pronounced splitting leads to substantial renormalizations of these levels and subsequently,  of the vertical excitation energies of the doubly degenerate singlet and triplet excited states. 
Our findings underscore the pressing need to incorporate quantum vibronic effects in first-principles calculations, particularly when comparing computed vertical excitation energies with experimental data.  
Our study also reveals the efficiency of stochastic thermal line sampling for studying phonon renormalizations of solid-state spin defects.
\end{abstract}

Spin defects in semiconductors \cite{RevModPhys2023} have the potential to realize quantum technologies working near room temperature.\cite{Acín_2018}. 
Various applications have been suggested in the literature, including quantum sensing,\cite{sensing1, sensing2} quantum communication,\cite{commun} and quantum computing.\cite{computing1, computing2}. However, despite rapid experimental and theoretical progress in the last decade, challenges remain in controlling and increasing the coherence time of spin qubits, which is ultimately limited by quantum vibronic effects (electron-phonon interactions).\cite{Wolfowicz2021}. Therefore, investigating electron-phonon interactions of spin defects in solids is essential to understand and improve their performance.  
     
Ground and excited state properties of solid-state spin defects, \textit{e.g.}, the nitrogen vacancy center (NV$^-$) in diamond, have been extensively studied using first principle methods, including density functional theory (DFT) and time-dependent DFT\cite{Gali_Rev_2019, Gali_PRL_2009, Gali_2011, Gali_PRB_2017, Alkauskas_2014, Razinkovas_PRB_2021, YuJin_PRM_2021, YuJin_NPJ_2022}, many-body perturbation theory (GW and Bethe Salpeter equation (BSE)\cite{Gali_PRB_2010}), quantum chemistry\cite{Bhandari_PRB_2021}, embedding methods\cite{Vorwerk2022, Nan_JCTC_2022, Haldar_JPCL_2023} and quantum Monte Carlo (QMC).\cite{Simula_PRB_2023} 
The predictions of the static vertical excitation energies (VEEs)  provided by these methods usually differ by several hundred meV,  as reported in Table-\ref{tab:static_vee}, and none of these predictions is in excellent agreement with experiments. It was speculated that such discrepancy may arise from neglecting  zero-point quantum vibronic effects of the ground state.\cite{Gali_PRL_2009} However, so far, no estimate of the effect of zero-point motion on VEEs of spin defects has been reported in the literature.

Based on the Allen-Heine-Cardona (AHC) theory,\cite{Allen_1976, Cardona_2005} first-principles calculations of the quantum vibronic effects in solids have been carried out in the literature with both perturbative \cite{Canuccia_PRL_2011, Ponce_2014, Ponce_2015, Giustino_Rev_2017, Han_JCTC_2022} or non-perturbative methods,\cite{Zacharias_PRL_2015, Zacharias_PRB_2016, TL_Monserrat_PRB_2016, Karsai_NJP_2018, Monserrat_Rev_2018} and all methods used, so far, only provided the phonon renormalizations of the single-particle electronic eigenenergies, not of the VEE between many-body excited states.  Perturbative methods employ (i) quantum harmonic oscillator (QHO) models for the potential energy surface (PES) and (ii) single-phonon and quadratic approximations for the single-particle electronic eigenenergies. The single-phonon approximation assumes that each phonon mode couples independently to the electronic levels (states) and multi-phonon effects are neglected. Within the single-phonon approximation, the quadratic approximation further assumes that the electronic eigenenergies vary quadratically as the ions are displaced from the equilibrium geometry along specific phonon modes (See section A1 in the Appendix for more details regarding these approximations). In addition, perturbative methods invoke the rigid-ion approximation, {\it i.e.} it is assumed that the ionic Hamiltonian only depends on the potential created by each nucleus independently. Although these approximations were shown to be valid for harmonic crystals like diamond \cite{Kundu_PRM_2021, Han_JCTC_2021, Han_JCTC_2022}, their applicability has been questioned for isolated molecules\cite{Gonze_2011, Kundu_PRM_2021}, molecular crystals\cite{Alvertis_PRB_2022, Kundu_JCTC_2023} and disordered solids\cite{Kundu_PNAS_2022}.

Alternatively, first-principles molecular dynamics (FPMD) simulations can be employed to sample ground and excited state energy surfaces, and these simulations do not utilize harmonic, independent-mode, and quadratic approximations. In principle, the energies of the many-body states could be obtained by carrying out advanced electronic structure calculations on sampled configurations.  However, FPMD treats the motion of the nuclei using classical mechanics, and its results are therefore only reliable above the Debye temperature. Although path-integral FPMD simulations provide an accurate framework to include nuclear quantum effects (NQE), they are rarely adopted to investigate quantum vibronic effects on electronic properties due to their  computational cost. Recently, we showed that by coupling a colored noise generalized Langevin equation thermostat, also known as quantum thermostat \cite{QT_Ceriotti_PRL_2009, QT_Ceriotti_JCTC_2010, QT_Review_Finocchi_2022} with FPMD simulation, it is possible to accurately predict NQEs on the electronic properties at approximately the same computational cost of  FPMD simulations.\cite{Kundu_PRM_2021} Nonetheless, these simulations remain computationally expensive when large supercells are employed, \textit{e.g.} those required to simulate solid-state spin-defects.\cite{Gali_Rev_2019} To lower the computational cost, one may compute ground, and excited state energies on sampled trajectories for a handful of configurational coordinates that define effective phonons; the electron-phonon problem is then solved in reduced dimensions along these specific configurational coordinates \cite{Gali_PRB_2017, YuJin_PRM_2021, YuJin_NPJ_2022} and as a consequence, the vibronic contribution of many of the phonon modes is neglected. 

In this work, taking the NV$^-$ center in diamond as an exemplary solid-state spin
defect, and using first principles calculations, we demonstrate the role of  zero-point quantum vibronic effects on the electronic properties of solid-state spin defects. In particular, we show that the
quantum vibronic effects substantially renormalize the single-particle electronic eigenenergies
and consequently, the vertical excitation energies computed from  many-
body excited states. Hence, it is essential to account for quantum vibronic effects when comparing the results of computations and experiments.  We also discuss  the impact of  nonquadratic electron-phonon coupling and multi-phonon processes on the dynamic Jahn-Teller effect (DJT)  that
has been observed experimentally in the NV$^-$ center in diamond.\cite{DJT_PRL_2009, DJT_PRL_2011} 

\begin{figure}[htbp]
\centering
\includegraphics[width=8.4cm]{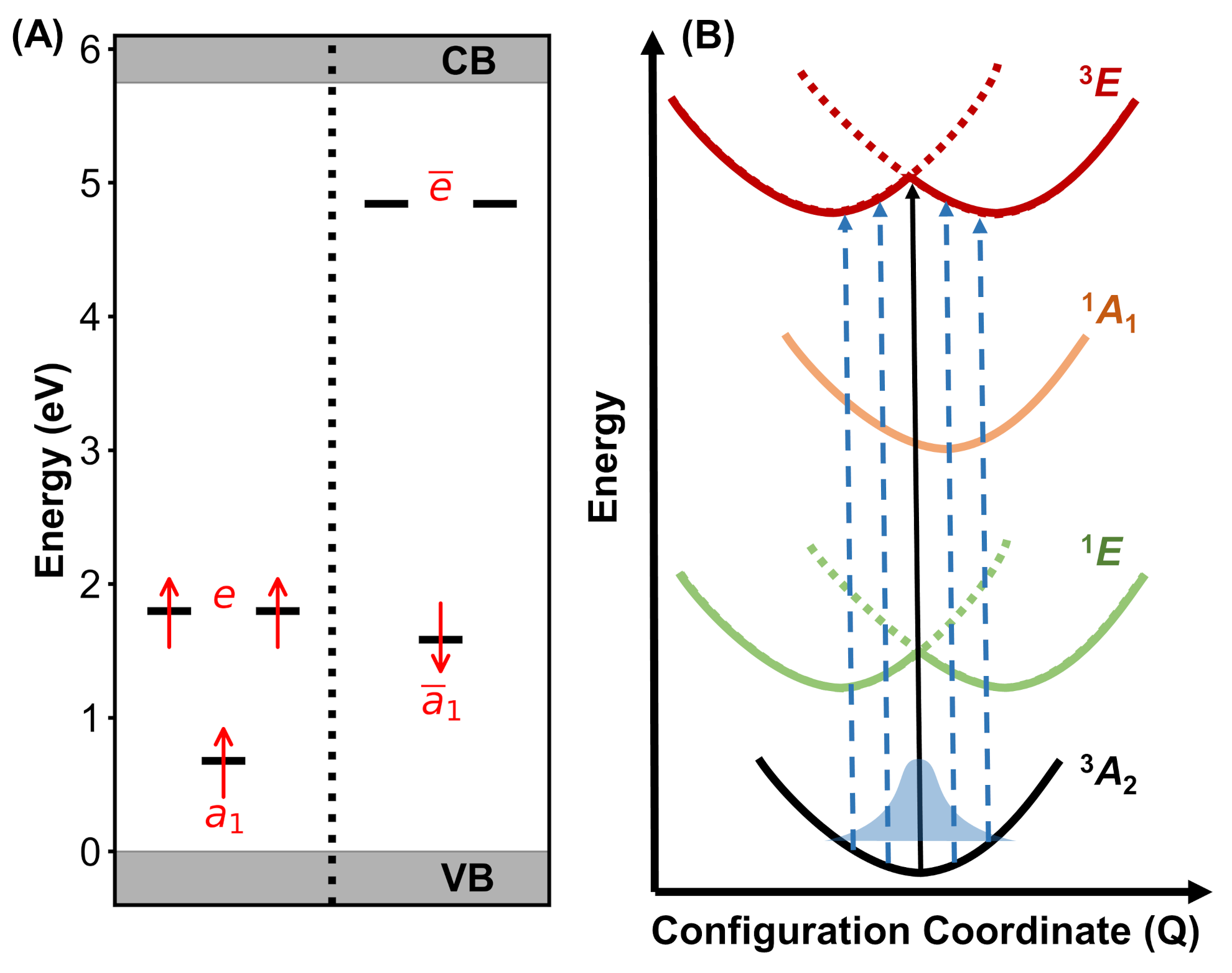}
\caption{(A) Energies of the single-particle defect levels (residing within the band gap of the diamond), obtained using spin-unrestricted density functional theory with the DDH hybrid functional. (B) Schematic representation of the potential energy surfaces of the many-body electronic states. Ground state vibronic effects drive the system out of the equilibrium geometry (obtained in the absence of the quantum vibronic effects), thus lowering the symmetry causing a dynamic Jahn-Teller splitting of the {\it E}-states. The dotted surfaces show the higher energy branch of the split {\it E} states. }
\label{fig:elec_struc}
\end{figure}

The geometrical configuration of the ground state of the NV$^-$ center in diamond belongs to the C\textsubscript{3v} point group and the electronic ground state has \textsuperscript{3}A\textsubscript{2} symmetry. The highest occupied and lowest unoccupied molecular orbitals (HOMO and LUMO, respectively) are the doubly degenerate $e$ levels. In addition to these levels, an occupied $a_1$ level resides within the band gap of diamond. In our calculations,  we used spin-unrestricted DFT, and we labeled  these defect levels in $\alpha$ and $\beta$ spin channels as $a_1$, $e$ and $\overline{a}_1$, $\overline{e}$ respectively. Fig\ref{fig:elec_struc}A shows the energies of these defect levels at the ground state equilibrium geometry as obtained with the dielectric-dependent hybrid (DDH)\cite{Skone_PRB_2014} functional. A spin-conserving electronic excitation: $\overline{a}_1 \rightarrow \overline{e}$ gives rise to the first triplet excited state $^3E$, while a linear combination of the spin-flip excitations: $a_1 \rightarrow \overline{e}$ and $e \rightarrow \overline{e}$ gives rise to  the low-lying singlet excited states: $^1A_1$ and $^1E$. 

\begin{table}[]
    \centering
    \caption{ Computed vertical excitation energies (VEE) (eV) of the NV$^-$ center in diamond using various theoretical approaches; time-dependent density functional theory (TDDFT) with semi-local (PBE) and hybrid (DDH) functionals, quantum defect embedding theory (QDET), density matrix embedding theory (DMET), many-body perturbation theory (GW+BSE, Bethe Salpeter Equation), quantum Monte Carlo (QMC), and the quantum chemistry method CASSCF. Experimental results, including those for the zero-phonon lines (ZPL) are also shown. See Fig. \ref{fig:elec_struc} for the labeling of states.}
    \begin{threeparttable}
    \begin{tabular}{cccc}
    \toprule
                       & $^1E$ & $^1A_1$ & $^3E$  \\
    \midrule      
    TDDFT (PBE)\cite{YuJin_TDDFT_2023}& 0.512&  1.336&  2.089\\
 TDDFT(DDH)\cite{YuJin_TDDFT_2023}& 0.681&  1.973&2.372\\
 QDET\cite{YuJin_TDDFT_2023}& 0.479& 1.317&2.162\\
 CASSCF\cite{Bhandari_PRB_2021}& 0.34& 1.41&1.93\\
 DMET\cite{Haldar_JPCL_2023}& 0.50& 1.56&2.31\\
 GW+BSE\cite{Gali_PRB_2010}& 0.40& 0.99&2.32\\
 QMC\cite{Simula_PRB_2023}& 0.81--1.06& 2.13--2.37&2.37--2.54\\
    Exp.\cite{Davies_1976, Rogers_NJP_2008, Goldman_PRL_2015, Kehayias_PRB_2013, Goldman_PRB_2015}                & 0.50 -- 0.59\tnote{a}& 1.76 -- 1.85\tnote{a}& 2.18\cite{Davies_1976} \\
    Exp. ZPL\cite{Davies_1976, Rogers_NJP_2008, Goldman_PRL_2015, Kehayias_PRB_2013, Goldman_PRB_2015}               & 0.34 -- 0.43\tnote{a}& 1.51 -- 1.60\tnote{a}& 1.94 \\
    \bottomrule
    \end{tabular}
    \begin{tablenotes}
        \item[a] Not directly measured; obtained from the measured ZPL/VEE's of $^3A_2 \rightarrow{}^3E$, $^1A_1 \rightarrow {}^1E$ and the energy difference between $^3E$ and $^1A_1$ states; see ref. \cite{Haldar_JPCL_2023}.
    \end{tablenotes}
    \end{threeparttable}
    \label{tab:static_vee}
\end{table}

Figure \ref{fig:elec_struc}B shows the potential energy surfaces (PESs) along a one-dimensional configuration coordinate for the excited states. Quantum vibronic effects due to ground state phonons drive the system out of the ground state equilibrium geometry determined in the absence of quantum motion, towards configurations with lower symmetry, thus causing a DJT splitting of the degenerate single-electronic levels ($e$,$\overline{e}$) and many-body states ($^1E$,$^3E$). In the presence of such splitting, we label the $e$ and $\overline{e}$ levels with lower and higher energy as $e_\mathrm{low}$, $\overline{e}_\mathrm{low}$ and $e_\mathrm{high}$, $\overline{e}_\mathrm{high}$, respectively; see Fig.\ref{fig:orbital-splitting}  in the Appendix A1. Similarly, we name the higher and lower branches of the doubly-degenerate many-body states as $^1E_\mathrm{low}$,$^3E_\mathrm{low}$ and $^1E_\mathrm{high}$, $^3E_\mathrm{high}$.   
 
To study the quantum vibronic effects on the electronic structure of the spin defect, we employed stochastic methods,\cite{Monserrat_Rev_2018} a non-perturbative technique that relies on the QHO approximation for the PESs but goes beyond the single-phonon and quadratic approximations. Models based on the QHO yield a Gaussian probability distribution for each phonon mode and the total probability distribution reduces to a product of independent Gaussian functions.\cite{Zacharias_PRL_2015,Monserrat_Rev_2018, Kundu_JCTC_2023} A Monte Carlo (MC) sampling is employed to sample each mode simultaneously, thus avoiding the need for the single-phonon approximation.\cite{Monserrat_Rev_2018} To accelerate the convergence of  the MC algorithm, Monserrat proposed to sample the thermal lines (TL) defined by two mean-value positions: $\sigma_{\nu,T}$ and $-\sigma_{\nu,T}$ for each phonon mode $\nu$, where $\sigma_{\nu,T}$ is the width of the resulting Gaussian function at temperature $T$.\cite{TL_Monserrat_PRB_2016} Later, Zacharias and Giustino proposed to consider  a specific TL defined by a set of special displacements (SD): $\{+\sigma_{1,T},-\sigma_{2,T},+\sigma_{3,T},...,(-1)^N\sigma_{N,T}\}$, where $N$ is the total number of phonon modes.\cite{Zacharias_PRB_2016, Zacharias_PRR_2020} The stochastic convergence for each of these methods can be improved significantly  using a variance reduction technique where antithetic pairs of each displaced configuration are included.\cite{Zacharias_PRB_2016, Zacharias_PRR_2020}. The stochastic (TL and MC) and SD methods are a few to several orders of magnitude faster than FPMD simulations.

 Various flavors of stochastic methods have been applied to semiconductor crystals,\cite{Karsai_NJP_2018}, perovskites \cite{Zacharias_NPJ_2023}, isolated molecules\cite{Kundu_JCTC_2023}, and molecular crystals. \cite{TL_Monserrat_PRB_2016, Alvertis_PRB_2022, Kundu_JCTC_2023} Despite recent efforts to include anharmonicity,\cite{Ning_PCCP_2022, Zacharias_NPJ_2023} the performance of these methods in a strongly anharmonic regime is not guaranteed.
For example,  we recently showed that  stochastic methods are applicable to molecules and molecular crystals composed of rigid molecules but introduce large errors in the case of  floppy molecules and their corresponding molecular crystals.\cite{Kundu_JCTC_2023}      

\begin{figure}[htbp]
\centering
\includegraphics[width=8.4cm]{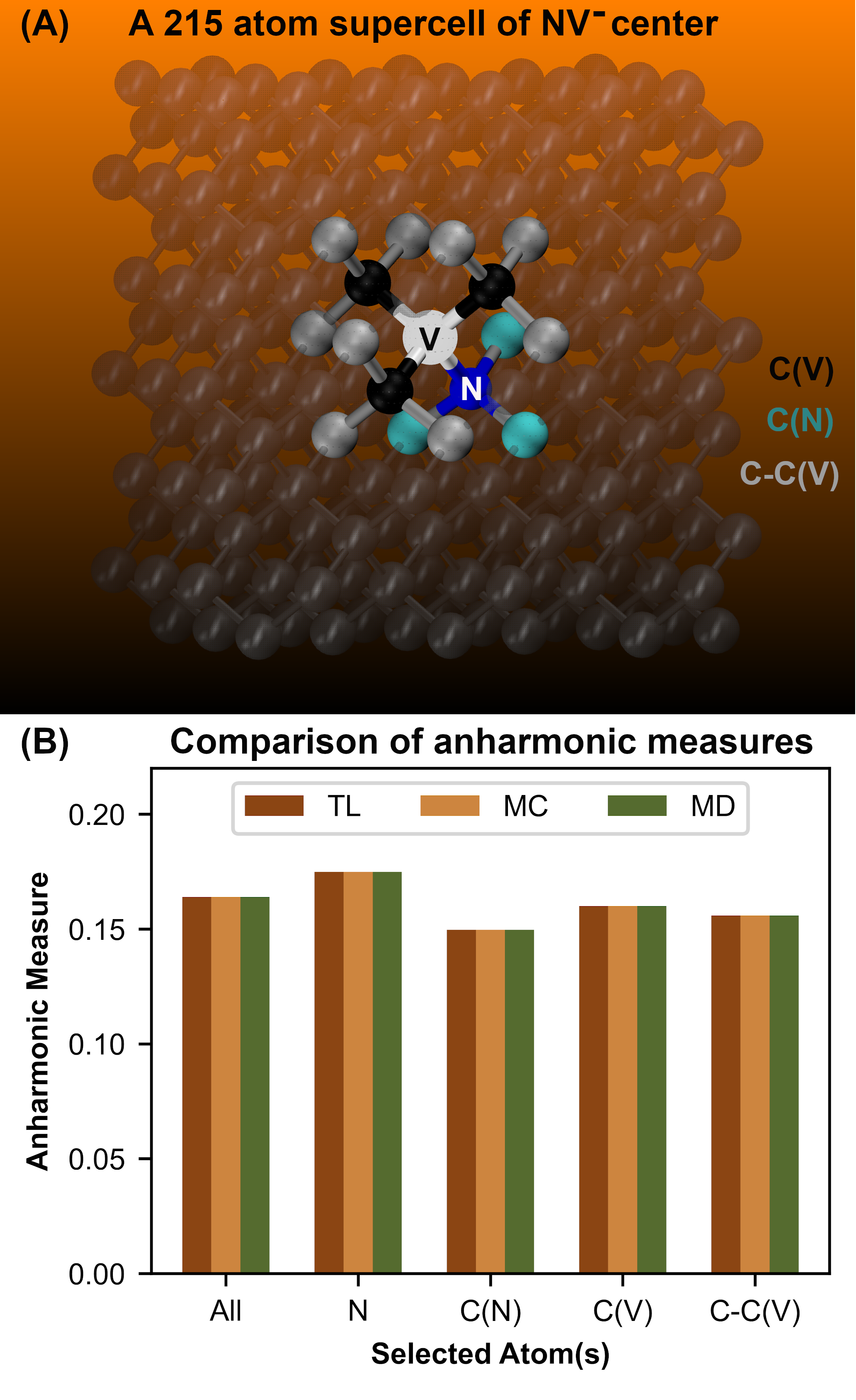}
\caption{(A) A C\textsubscript{214}N supercell model of the NV$^-$ center in diamond highlighting the position of the vacancy (white sphere), V, nitrogen (blue sphere), N, first carbon (cyan spheres) neighbors to the N, C(N), and the first (black spheres) and second carbon (grey sphere) neighbors to the vacancy, C(V) and C-C(V), respectively. (B) Anharmonic measures (see text) at 300 K computed with the PBE functional for the C\textsubscript{214}N supercell using the configurations obtained from thermal line (TL) sampling,  Monte Carlo (MC) sampling, and molecular dynamics(MD) with a quantum thermostat. }
\label{fig:anh_mes}
\end{figure}

To examine the performance of stochastic methods to accurately compute quantum vibronic effects on the electronic properties of the NV$^-$ center, we performed a thermal line (TL) and a Monte Carlo (MC) simulation with a $3\cross3\cross3$  supercell (C\textsubscript{214}N, see Fig\ref{fig:anh_mes}A). We then compared the results with those of FPMD simulations coupled to a quantum thermostat. All calculations were carried out at 300 K with the generalized gradient approximated (GGA) Perdew-Burke-Erzenhof (PBE) functional, \cite{PBE_Perdew_PRL_1996_1, PBE_Perdew_PRL_1996_2}  which yields results for diamond vibrational density of states in agreement with those of  hybrid functionals\cite{YuJin_PRM_2021}. FPMD simulations were carried out  using the i-PI---Qbox coupling scheme \cite{Kundu_PRM_2021, ipi-qb-coupling}, where the i-PI package \cite{ipi_Kapil_2018} moves the nuclei and the Qbox code \cite{Qbox_Gygi_2008} provides forces on the nuclei by solving the Kohn-Sham equations. For the stochastic simulations, we generated 200 independent displaced configurations and their antithetic pairs (resulting in 400 configurations) using the PyEPFD package \cite{Kundu_PRM_2021, Kundu_JCTC_2023, pyepfd} and performed DFT single-point calculations using the Qbox code (see section A3 in the Appendix for further computational details).

We compared the anharmonic measure proposed by Knoop {\it et al}, \cite{Knoop_anh_mes_PRM_2020}
\begin{equation*}
    a_x(T) =\sqrt{\frac{\sum_{x \in \bfA}\langle (F_x-F_x^\text{H})^2 \rangle_T}{\sum_{x \in \bfA}\langle F_x^2 \rangle_T}},
\end{equation*}
as obtained from stochastic calculations and FPMD simulations.  For a specific configuration, $x$,  $F_x$ and $F_x^\text{H}$ denote the forces obtained from a DFT calculation and from the harmonic approximation, respectively; $\bfA$ denotes the subspace of a set of chosen atoms, and $\langle.\rangle_T$ denotes the ensemble average. Fig. \ref{fig:anh_mes}A highlights the different sets of chosen atoms, while the anharmonic measures for different ensembles are summarized in Fig. \ref{fig:anh_mes}B. Small ($<0.2$) but non-zero anharmonic measures for various atom sets indicate that the system is weakly anharmonic, and identical anharmonic measures obtained with stochastic calculations and MD simulations show that the former can accurately capture the weak anharmonicities of the NV$^-$ center in diamond.    

\begin{figure}[htb]
\centering
\includegraphics[width=8.4cm]{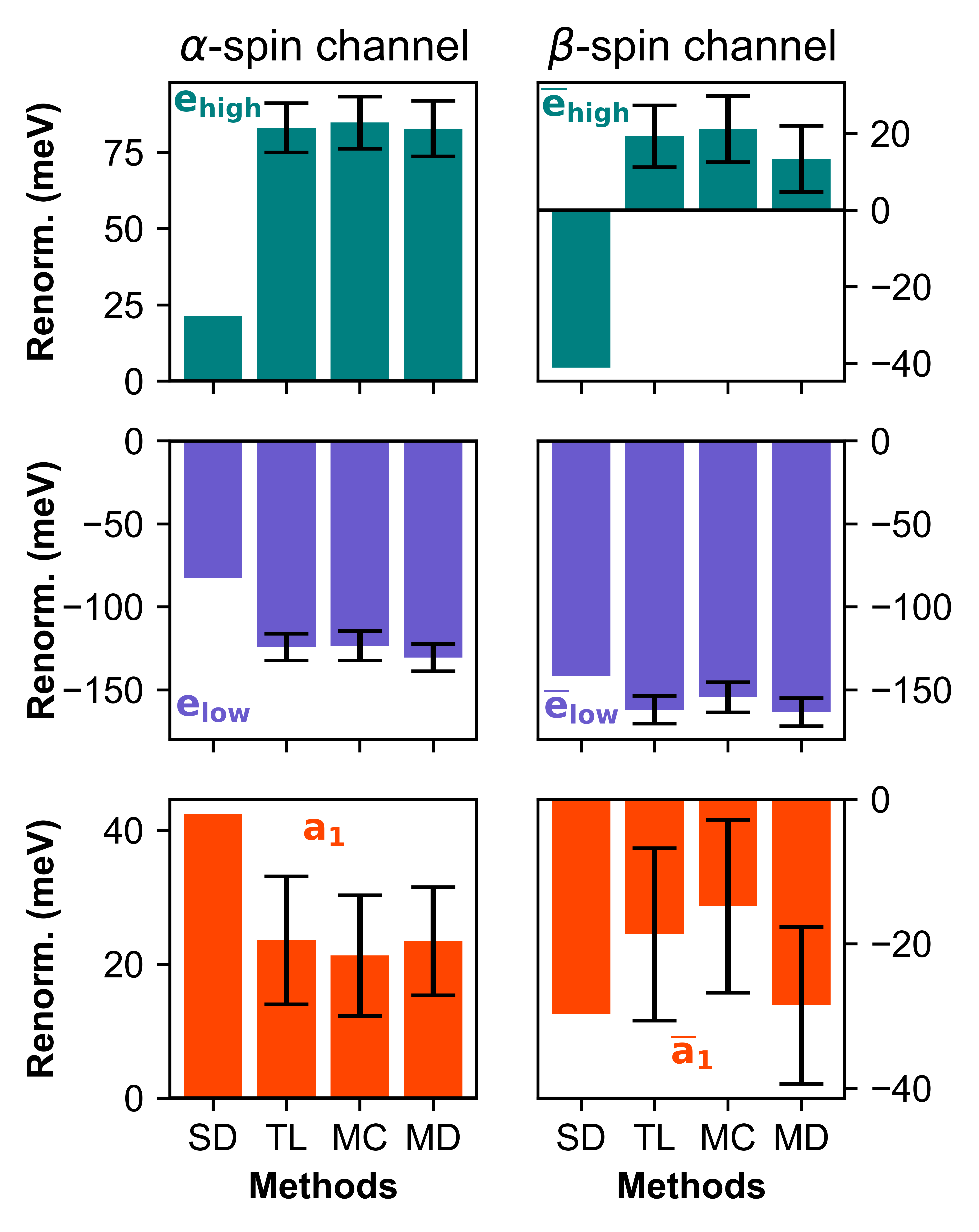}
\caption{The ground state phonon renormalizations of the single-particle defect levels ($a_1$, $\overline{a}_1$, $e_{\mathrm{low}}$, $e_{\mathrm{high}}$, $\overline{e}_{\mathrm{low}}$, $\overline{e}_{\mathrm{high}}$) at 300 K obtained with various sampling techniques: special displacement (SD), thermal line (TL), Monte Carlo (MC), and first principles molecular dynamics (MD) with a quantum thermostat. For these calculations, we used the PBE functional and a C\textsubscript{214}N supercell.Black bars indicate the estimated error bars. See also Fig. \ref{fig:orbital-splitting} in the Appendix for the nomenclature of these levels.}
\label{fig:methods_compare}
\end{figure}

Fig. \ref{fig:methods_compare} compares the phonon renormalizations of the single-particle defect level energies obtained with stochastic and  FPMD simulations. While $e$ and $\overline{e}$ levels split into lower ($e_{\mathrm{low}}$, $\overline{e}_{\mathrm{low}}$) and higher  ($e_{\mathrm{high}}$, $\overline{e}_{\mathrm{high}}$) branches with a substantial ($>100$ meV) renormalization in the lower branches, $a_1$ and $\overline{a}_1$ levels show a smaller (20--40 meV) phonon renormalizations. Notably, both TL and MC sampling yield phonon renormalizations comparable to that of FPMD simulations, showing the accuracy of stochastic calculations.   

We also carried out calculations  with the SD method whose computational workload is roughly two orders of magnitude smaller than that of the TL(MC) sampling,  as it requires only 2 DFT calculations (when antithetic pairs are used) instead of a few hundred. The renormalizations of the energy levels obtained with the SD method, unfortunately, are not adequate for all defect levels, especially for the $\overline{e}_{\mathrm{high}}$ level, the method predicts a negative value, while the other stochastic methods and our reference MD simulation predict positive values. Further, the SD method predicts only a 101 meV difference between $\overline{e}_{\mathrm{high}}$ and $\overline{e}_{\mathrm{low}}$ levels, a measure of DJT splitting energy for the $\overline{e}$--levels, compared to our reference value of 177 meV obtained using FPMD simulation. Below we show that this energy difference is directly related to the splitting measure of the many-body $^3E$ state. 

We note that the SD method is only guaranteed to be accurate in the limit of large supercells.\cite{Zacharias_PRB_2016, Zacharias_PRR_2020}. In the case of the delocalized band edges, the band gap obtained with the SD method is within 2.5\% of the TL(MC) results even for a C\textsubscript{214}N supercell and the agreement improves for a larger (C\textsubscript{510}N) supercell, see Table-\ref{tab:band_gap} in Appendix A4.
However, for the renormalizations for the localized defect levels, even for a C\textsubscript{510}N supercell, the SD results are not yet converged to those of the TL(MC) method, see Fig. \ref{fig:methods_compare_nv512} in the Appendix. We conclude that when using a moderate-sized supercell (C\textsubscript{214}N or C\textsubscript{510}N), calculations considering only a special displacement are not adequate to yield converged results for the phonon renormalizations of all defect levels, and ultimately of the VEEs of the many-body states.  Therefore, the SD method will not be used any further in this study, and we will only adopt the TL sampling method.    

Now we turn to discussing finite size effects on the computed phonon renormalizations of the defect levels. We performed additional TL sampling simulations for $n \cross n \cross n$ supercells with $n=2$ (C\textsubscript{62}N) and $n=4$ (C\textsubscript{510}N) using the PBE functional and we extrapolated our results to the thermodynamic limit using $\Delta E_n = \Delta E_{\infty}+B/n^3$  (see Fig. \ref{fig:size_func_compare}A and section A5 in the Appendix). We notice that as the supercell size increases, the renormalizations of the lower branch of the $e$ and $\overline{e}$ levels increase, causing a larger splitting of the $e$ and $\overline{e}$ levels. Nevertheless, the trend obtained with the smallest supercell is correct and the results obtained with the C\textsubscript{214}N supercell are practically converged, justifying the use of the 215 atom supercell for the study of  phonon renormalizations of the NV$^-$ center with more accurate electronic structure theories,  including hybrid DFT.    

\begin{figure}[tbh]
\centering
\includegraphics[width=8.4cm]{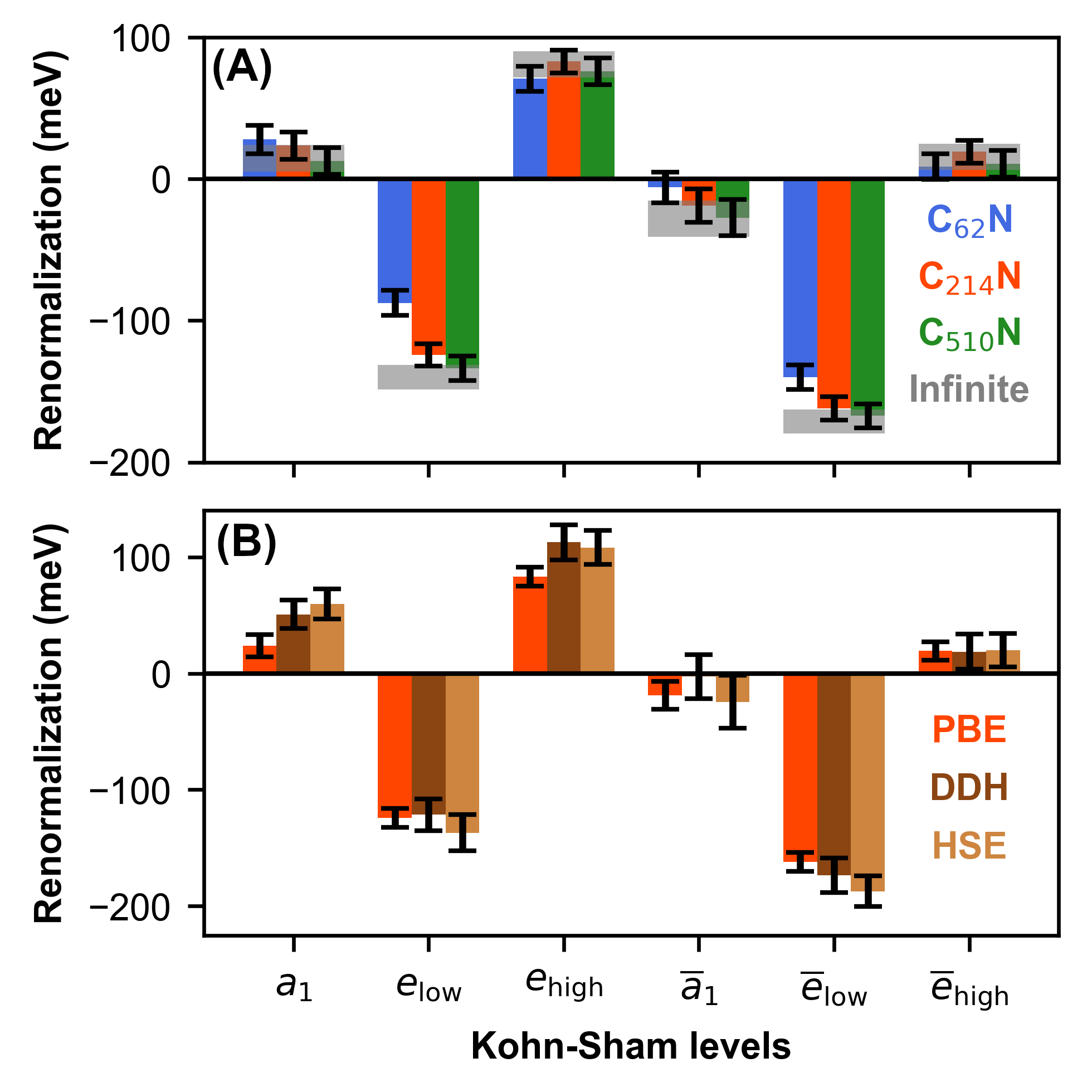}
\caption{The ground state phonon renormalizations of the single-particle defect level energies at 300 K as obtained employing (A) the PBE functional but with different supercells and (B) different functionals (PBE and two hybrid functionals DDH and HSE) with a C\textsubscript{214}N supercell. The gray-shaded region in panel A denotes the results extrapolated to the thermodynamic limit. See Table \ref{tab:fss_eigval_renorm} in Appendix A5 for the numerical data. Black bars indicate the estimated error bars. See also Fig.  in the SI for the nomenclature of these levels.}
\label{fig:size_func_compare}
\end{figure}

We then proceeded to carry out  TL samplings at 300 K using a C\textsubscript{214}N supercell  and compared the results obtained with two hybrid functionals: DDH \cite{Skone_PRB_2014} and HSE\cite{hse1,hse2}, as shown  in Fig. \ref{fig:size_func_compare}B.
To reduce the computational cost, we generated only 100 independent TL samples and their antithetic pairs which yielded  reasonably converged phonon renormalizations (see section A3 in the Appendix). Note the similarities between DDH and HSE results indicating that range separation has a minor  effect on the phonon renormalizations. In addition, both GGA and hybrid functionals predict a similar trend for the renormalizations, though hybrid calculations predict slightly larger values for both the renormalizations (for the $a_1$, $e$, and $\overline{e}$ levels) and the DJT splitting energies of the degenerate levels (25--30 and 10--40 meV larger, respectively). 

\begin{table}[htbp]
    \centering
    \caption{Comparison of phonon renormalizations of single-particle defect level energies (meV) at 0 K computed with different methods: stochastic thermal line sampling (TL), frozen phonon (FP), and density matrix perturbation theory (DMPT). The results are shown for a C\textsubscript{62}N supercell with PBE and DDH hybrid functionals.See also Fig.\ref{fig:orbital-splitting} in the Appendix for the nomenclature of these levels.}
    \begin{threeparttable}
    \begin{tabular}{cccccccccc}
    \toprule
       Functional & Method & $a_1$ & \multicolumn{3}{c}{$e$} & $\overline{a}_1$ & \multicolumn{3}{c}{$\overline{e}$} \\
       \cmidrule(lr){4-6} \cmidrule(lr){8-10}
       & & & $e_{\mathrm{low}}$ & $e_{\mathrm{high}}$ & $e_{\mathrm{avg}}$ & & $\overline{e}_{\mathrm{low}}$ & $\overline{e}_{\mathrm{high}}$ & $\overline{e}_{\mathrm{avg}}$ \\
    \midrule
       PBE & TL\tnote{a} & 24 & -91 & 68 & -12 & -10 & -141 & 6  & -67 \\
           & TL-sp& 23 & -46 & 24 & -11 & -13 & -130 & -3 & -67 \\
           & FP & 25 & -13 & -11 & -12 & -12 & -69 & -65 & -67 \\
           & DMPT\cite{Han_JCTC_2022}& 38 & - & - & 1 & 12 & - & - & -35 \\
    \hline
       DDH & TL\tnote{a} & 44 & -83 & 101 & 9 & -7 & -151 & 15 & -67 \\
           & FP & 43 & 4 & 3 & 4 & -6 & -63 & -67 & -65 \\
           & DMPT\cite{Han_JCTC_2022} & 52 & - & - & 9 & 8 & - & - & -33 \\
     \bottomrule
     \end{tabular}
     \begin{tablenotes}
        \item[a] Estimated stochastic uncertainties are 6-10 meV based on 400 sampled configurations.
     \end{tablenotes}
    \end{threeparttable}
    \label{tab:methods}
\end{table}

To better understand the mechanism of the DJT splitting, we performed additional finite-difference frozen-phonon (FP) calculations at 0 K for the C\textsubscript{62}N supercell with PBE and DDH functionals. We compared  our FP results with those of DMPT\cite{Han_JCTC_2022} and TL sampling results in Table \ref{tab:methods}. Note that being based on
non-degenerate perturbation theory, DMPT does not predict a finite  DJT splitting of degenerate levels.

For the non-degenerate $a_1$ and $\overline{a}_1$ levels, all methods yield consistent results: slightly positive renormalizations for the former while almost negligible renormalizations for the latter. For the degenerate levels $e$ and $\overline{e}$, however, only the average renormalizations of their lower and higher branches, when calculated using TL and FP methods, should be compared with the DMPT results. Although  for the $e$ level, the average renormalization is negligibly small with all methods, there is a considerable difference between DMPT and FP(TL) results for that of the $\overline{e}$ level. We recall that DMPT employs the rigid ion approximation which is at the origin of such discrepancies for the localized defect levels of the NV$^-$ center in diamond. 

Interestingly, FP  also predicts negligibly small (1--4 meV) splitting of the $e$ and $\overline{e}$ levels. We recall that FP (DMPT) employs single-phonon and quadratic approximations which likely underestimate the DJT splitting. To disentangle the impact of these approximations, we applied TL sampling along each phonon mode separately thus only single-phonon (sp) processes are considered but the quadratic approximation is not employed. We refer to this approach as "TL-sp" and for the PBE functional, we compare the FP and TL results in Table \ref{tab:methods}. The  "TL-sp" and TL approaches yield almost the same splitting for the $\overline{e}$ level. This indicates that for the unoccupied $\overline{e}$ levels, the single-phonon approximation is reasonable, and multi-phonon effects are negligible,  and the DJT  splitting mainly stems from the non-quadratic electron-phonon coupling. In contrast, for the occupied $e$ levels, "TL-sp" and TL results differ, indicating  the importance of both non-quadratic electron-phonon coupling and multi-phonon effects in determining the value of the DJT splitting energy.              

Having understood the zero phonon renormalization of single-particle energy levels, we now turn to consider  the VEEs from the $^3A_2$ ground state to many-body excited states, starting from  $^3A_2 \rightarrow {}^3E$. Previous studies have shown that the wavefunction of the $^3E$ state can be represented by a single Slater determinant obtained with a neutral excitation of an electron from $\overline{a}_1$ to $\overline{e}$ level.\cite{YuJin_PRM_2021}
Neglecting electron-phonon interactions, vertical excitation energies have been computed  previously at the $^3A_2$ ground state geometry by constrained occupation DFT (CDFT),\cite{Gali_PRL_2009, Razinkovas_PRB_2021,YuJin_PRM_2021} TDDFT, \cite{Gali_2011, YuJin_NPJ_2022, YuJin_TDDFT_2023} and  quantum  embedding theories \cite{Vorwerk2022, Nan_JCTC_2022, Haldar_JPCL_2023} or QMC\cite{Simula_PRB_2023}.
As expected, many body theories  predict vertical excitations different from those of single particle energy (Kohn-Sham eigenvalues) differences. However, it is interesting to understand whether  an estimate of the VEE renormalizations can be obtained just from the difference in phonon renormalizations of the single-particle energies. The ability to obtain  this estimate would result in considerable computational savings, compared to sampling different configurations and then  performing electronic structure calculations with high-level theories on each of these configurations.

\begin{figure}[htbp]
\centering
\includegraphics[width=8.4cm]{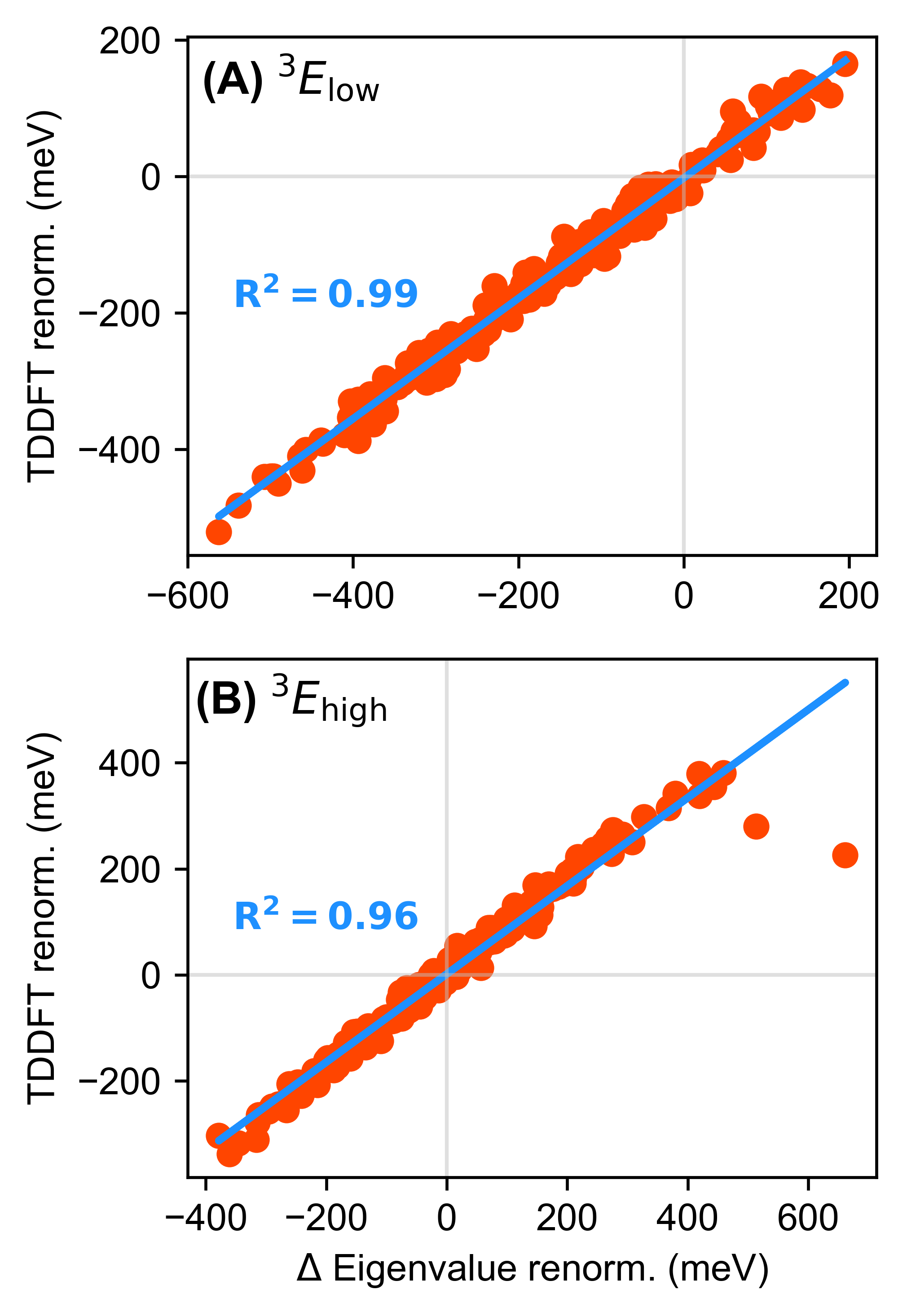}
\caption{The correlation between the renormalizations of the vertical excitation energies at 300 K computed using (i) time-dependent density functional theory (TDDFT) and (ii) Kohn-Sham eigenvalue renormalization difference for the (A) lower and (B) higher branches of the \textsuperscript{3}{\it E} state. Thermal line sampling was employed to generate 200 displaced configurations for a C\textsubscript{214}N supercell and the DDH hybrid functional was used to compute the renormalizations for each displaced configuration}
\label{fig:renorm_correlation}
\end{figure}

We performed TDDFT/DDH calculations with a C\textsubscript{214}N supercell for the $^3E$ state for the same 200 configurations used to compute the single-particle renormalizations. We used the TDDFT implementation \cite{YuJin_TDDFT_2023} of the WEST code\cite{west1, west2} after obtaining the ground state wavefunctions with the Quantum Espresso package\cite{qe1, qe2}. Due to the DJT effect, the $^3E$ state splits into $^3E_{\mathrm{low}}$ and $^3E_{\mathrm{high}}$ branches. These two branches can be thought of as originating from the following single-particle electronic excitations: $i \rightarrow f$ with $i=\overline{a}_1$, $f=\overline{e}_{\mathrm{low}}$ and $\overline{e}_{\mathrm{high}}$, respectively. Therefore VEE renormalizations can be estimated from the difference: $\Delta E_f - \Delta E_i$, where $\Delta E$ denotes the phonon renormalizations of the Kohn-Sham eigenvalues of the respective levels, $f$ and $i$. In Fig. \ref{fig:renorm_correlation} we present a correlation plot of the VEE renormalizations obtained from (i) TDDFT, and (ii) Kohn-sham eigenvalue renormalization difference for the 200 sampled configurations. 

For both the lower and higher branches of $^3E$, TDDFT renormalizations not only show a near-perfect linear correlation with the Kohn-Sham eigenvalue renormalization differences but also the differences in their expectation values, 18 and 2 meV, for $^3E_\mathrm{low}$ and $^3E_\mathrm{high}$ branches, respectively, are smaller than the estimated uncertainties (21--25 meV). 
This clearly demonstrates that if an excited state is not multi-configurational, not only its VEE renormalization but also the value of its DJT splitting energy (if present) can be predicted just from the energy renormalizations of the single-particle levels, due to the similar effect of defect level relaxations (upon single-particle electronic excitation) in the optimized geometry and in a displaced configuration.  

Instead, the VEEs for the low-lying singlet states ($^1E$,$^1A_1$) which are multi-configurational, may not be obtained from single-particle renormalization differences, as we discuss below. We used the DDH functional and the spin-flip TDDFT implementation \cite{YuJin_TDDFT_2023} of the WEST package \cite{west1, west2}, which was  shown to be adequate to reproduce the VEEs of the low-lying singlet states of the NV$^-$ center.\cite{YuJin_NPJ_2022}. We carried out spin-flip TDDFT calculations for the 200 and 400 configurations obtained from TL sampling at 300 K  in the C\textsubscript{214}N and C\textsubscript{62}N supercells, respectively. Due to the DJT effect, the doubly-degenerate $^1E$ state splits into a lower ($^1E_\mathrm{low}$) and a higher ($^1E_\mathrm{high}$) branch. 

\begin{table}[]
    \centering
    \caption{ Static and phonon renormalized vertical excitation energies (VEEs) (eV), from the $^3A_2$ ground state to the states listed in the table, computed using time-dependent density functional theory (TDDFT) with the DDH functional, compared with the experimental values. The phonon renormalizations are computed using 200 sample configurations generated employing thermal line sampling (standard errors are shown).}
    \begin{threeparttable}
    \begin{tabular}{cccccc}
    \toprule
                       & $^1E_{\mathrm{low}}$ & $^1E_{\mathrm{high}}$ & $^1A_1$ & $^3E_{\mathrm{low}}$ & $^3E_{\mathrm{high}}$  \\
    \midrule      
    Static VEE\tnote{a}    & 0.68& & 1.96 & 2.40& \\
    Renorm. VEE \tnote{a}  & 0.53 $\pm$ 0.01  & 0.76 $\pm$ 0.01 & 1.94 $\pm$ 0.01  & 2.24 $\pm$ 0.02  & 2.42 $\pm$ 0.02 \\
    Exp. VEE               & 0.50 -- 0.59\tnote{b}& & 1.76 -- 1.85\tnote{b} & 2.18\cite{Davies_1976}& \\
    \bottomrule
    \end{tabular}
    \begin{tablenotes}
        \item[a] Extrapolated to infinite supercell limit based on the calculations on C\textsubscript{62}N and C\textsubscript{214}N supercells; see Table \ref{tab:tddft_all}. 
        \item[b] Not directly measured; obtained from the measured ZPL/VEE's of $^3A_2 \rightarrow{}^3E$,\newline $^1A_1 \rightarrow {}^1E$ and the energy difference between $^3E$ and $^1A_1$ states; see ref. \cite{Haldar_JPCL_2023}.
    \end{tablenotes}
    \end{threeparttable}
    \label{tab:excited_renorm}
\end{table}

While the phonon renormalization  of the $^1A_1$ state is marginal (-21 meV), it is substantial, -150 and -160 meV, for the lower branches of the doubly degenerate states, $^1E_\mathrm{low}$ and $^3E_\mathrm{low}$, respectively. The higher branch of the $^1E$ state also shows a considerable phonon renormalization (85 meV) leading to  a DJT splitting energy of 235 meV. The higher branch of the $^3E$ state, however, shows only a 20 meV phonon renormalization and  the corresponding DJT splitting energy  is 180 meV. The different VEE renormalizations for the various excited states highlight the complex nature of the excited state PES, where each state couples with a different set of phonons. This suggests that the displaced harmonic oscillator model or  the construction of the PES using configurational coordinates, as frequently done to study excited states of solids,\cite{Gali_Rev_2019, Gali_PRL_2009, Gali_PRB_2017, YuJin_PRM_2021, YuJin_NPJ_2022} may not be adequate in some cases. 

Although it is possible to obtain the contributions of each spin-flip excited Slater determinant to the singlet VEE renormalization, the coefficients in front of these contributions are not known a priori, before performing a full TDDFT calculation.
Hence VEE renormalizations of the singlet states cannot be predicted from the renormalization of the single particle orbitals. The only conclusion that can be drawn is that if the single-particle renormalizations of the defect levels involved in the spin-flip excitation are negligible then VEE renormalizations would be negligible too but its converse is of course not true. 

Table \ref{tab:excited_renorm} compares the phonon-renormalized VEEs with those obtained from experiments. We note that being spin-forbidden, VEEs of the singlet excited states from the triplet ground state cannot be measured directly but they can be deduced from the experimental zero-phonon lines (ZPLs) and VEEs for the $^3A_2 \rightarrow {}^1E$ and  $^1A_1 \rightarrow {}^1E$ transitions and the difference in energy between $^3E$ and $^1A_1$ states.\cite{Haldar_JPCL_2023} The computed VEEs for the lower branch of both the $^1E$ and $^3E$ states agree remarkably well with those obtained from experiments;  a slight (0.06 eV) difference for the latter is acceptable, considering that TDDFT only includes excited singles Slater determinants. Our results show that is it crucial to include quantum vibronic effects on energy levels when comparing to experiments. 

\begin{figure}[htbp]
\centering
\includegraphics[width=8.4cm]{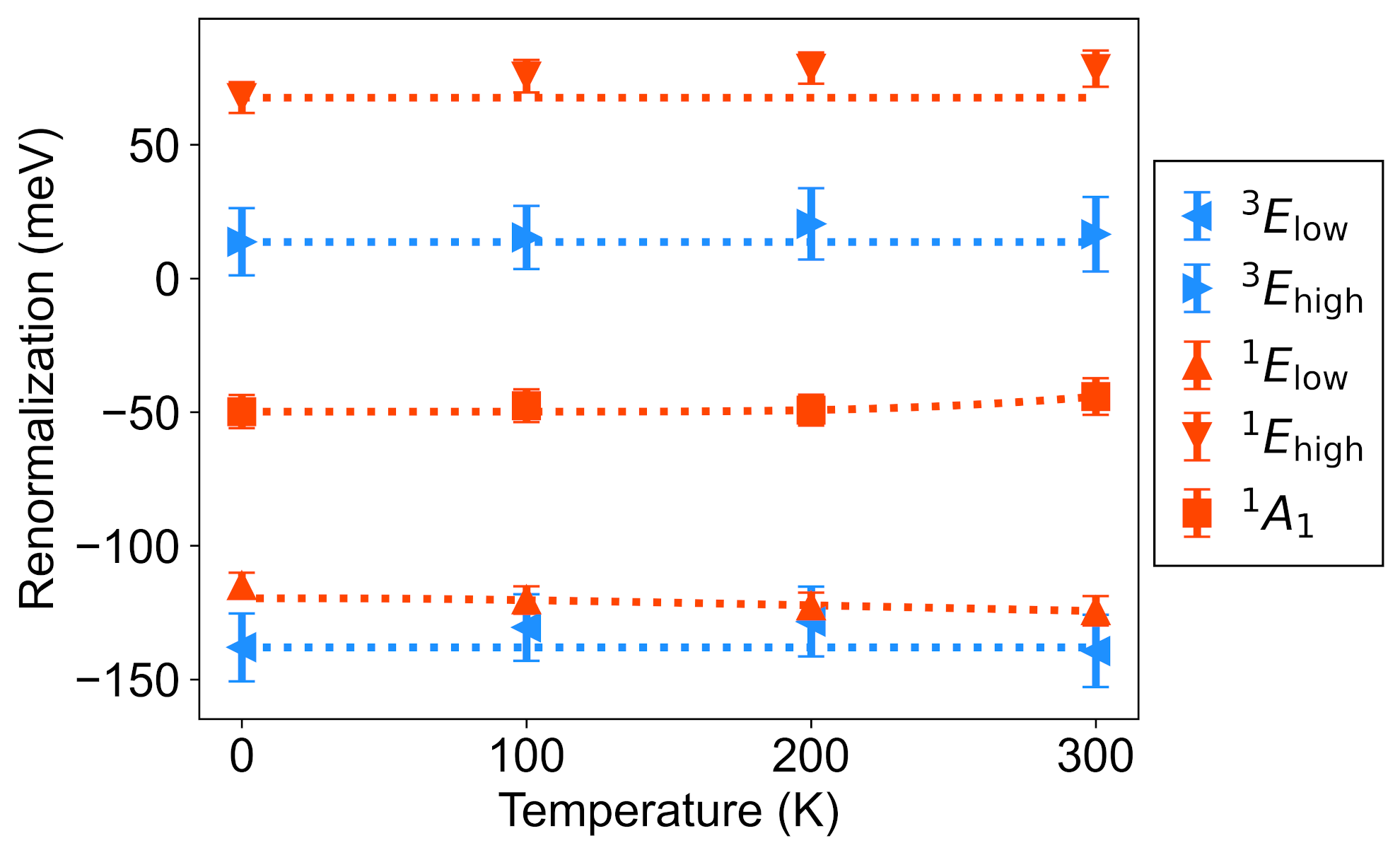}
\caption{Temperature dependence of the renormalization of the vertical excitation energies of the various excited states from the ground state, as obtained with thermal line sampling for a C\textsubscript{62}N supercell using the DDH functional. The symbols and lines represent the simulation data and the Vi\~{n}a model\cite{Vina_PRB_1984} fit, respectively.}
\label{fig:temp_effects}
\end{figure}

Finally, we turn to discuss the temperature ($T$) effects on the renormalizations of the single-particle energies as well as the VEEs for the many-body excited states. For this purpose, we performed TL samplings with DFT and TDDFT using the PBE and DDH functionals for a C\textsubscript{62}N supercell at $T = 0, 100, 200, 300$ K. For these temperatures, we also performed TL samplings with the PBE functional for a larger supercell (C\textsubscript{510}N). Fig. \ref{fig:temp_effects} shows only the VEE renormalizations obtained with the DDH functional for the C\textsubscript{62}N supercell as a function of $T$ (see section A6 in the Appendix for comparisons of the temperature dependence of phonon renormalizations of  defect levels and VEEs obtained with different functionals and supercells). Within 0--300 K, the $T$-dependence of chosen single-particle defect levels and VEEs of many-body excited states is marginal, as expected, since  the $T$-range considered here is way below the Debye temperature of diamond ($\simeq 2000$ K) and the zero-point vibrational motion represents the major contribution to the phonon renormalization of the electronic states. Thus, the normalized VEEs provided in Table \ref{tab:excited_renorm} are accurate within 0--300 K.   

In summary, taking the NV$^-$ center in diamond as an example, we showed that for a crystalline defect with light constituent atoms (high Debye $T$), even below 300 K, the energy expectation values for the electronic levels (single-particle) and many-body states are significantly renormalized (on the order of hundred meV) due to quantum vibronic effects of the ground state. Our findings point to the need for including such vibronic effects when comparing computed vertical excitation energies with those obtained from experiments.
Furthermore, our calculations showed that electronic states (levels) are rather insensitive to thermal fluctuations within 0--300 K, a desirable property for the use  of spin defects in room temperature quantum technologies.

We found that the zero-point vibrational  motion leads to the dynamic Jahn-Teller splitting of the degenerate electronic states (levels), causing renormalizations for these states (levels). 
Our calculations further revealed that the dynamic Jahn-Teller effect stems from both non-quadratic electron-phonon coupling and multi-phonon processes and therefore, it can not be straightforwardly described  using a perturbative or frozen phonon (finite difference) method.  Nonperturbative stochastic methods,  computationally cheaper alternatives to path-integral first-principles molecular dynamics, represent a promising approach to include non-quadratic and multi-phonon effects  when studying point defects. However, our study also suggests that when moderate-sized (n × n × n, with n = 3,4) supercells are used, the special displacement method (at least, in its original real-space implementation\cite{Zacharias_PRB_2016}), which is computationally efficient, is, unfortunately, inadequate to compute the renormalization energies of spin defects in diamond but the thermal line sampling method is accurate even for such supercells.  We have also shown that if the ground and excited states are not multi-configurational, the vertical excitation energy renormalizations can be reliably obtained from the energy renormalizations of the single-particle levels. Future work will address vibronic effects on other spin defects in diamond and silicon carbide and on emission energies.

\newpage

\section{Appendix}

\subsection{A1. Single-Phonon and Quadratic  Approximations}

A Quantum Harmonic Oscillator yields a Gaussian probability distribution\cite{Kundu_JCTC_2023} : 
\begin{equation}
    G(X_\nu;\sigma_{\nu,T}) = \frac{1}{\sqrt{2\pi\sigma_{\nu,T}^2}}\exp\qty(-\frac{X_{\nu}^2}{2\sigma_{\nu,T}^2})
\end{equation}
for each phonon mode $(X_{\nu})$ and the total probability distribution, $W(\bfX,T)$, reduces to a product of independent Gaussian functions with widths $\sigma$ related to the Bose occupation factor $n_B$: 
\begin{equation}\label{eq:sigma}
    \sigma_{\nu,T}= \sqrt{\frac{2n_B\qty(\omega_\nu,T)+1}{2\omega_\nu}}.
\end{equation}
where $T$ is the temperature. Therefore, the phonon-renormalized electronic observable, $\aobs(\bfX)$, is given by,
\begin{equation}\label{eq:ha_avg}
    \langle \aobs \rangle_T = \int d\bfX W(\bfX,T) \aobs(\bfX) = \int d\bfX \qty[\prod_{\nu} G(X_\nu;\sigma_{\nu,T})]\aobs(\bfX) 
\end{equation}
We note that though Eq. \ref{eq:ha_avg} is valid under the harmonic approximation, it does not assume any explicit dependence of the electronic observable $\aobs$ on phonon modes, $\bfX=({X_1,X_2, ..., X_{\nu},...})$. A Monte Carlo (MC) or thermal line (TL) sampling directly evaluates this expression.

To further simplify the expression, we Taylor expand  $\aobs(\bfX)$:
\begin{equation}\label{eq:obs_taylor}
    \aobs(\bfX) = \aobs(\bfzero) + \sum_{\nu,i}\aobs^i_{\nu}X_{\nu}^i + \sum_{\nu<\nu',i,j} \aobs^{i,j}_{\nu,\nu'}X_{\nu}^iX_{\nu'}^j + ...
\end{equation}
where
\begin{equation}\label{eq:pdvs}
    \aobs^{i}_{\nu} = \frac{\partial^{i}}{\partial X_{\nu}^{i}} \aobs(\bfzero), \quad  
    \aobs^{i,j}_{\nu,\nu'} = \frac{\partial^{i}}{\partial X_{\nu}^{i}}\frac{\partial^{j}}{\partial X_{\nu'}^{j}} \aobs(\bfzero), \quad ...
\end{equation}
After inserting the expression from \ref{eq:obs_taylor} into Eq.\ref{eq:ha_avg}, we obtain,
\begin{equation}\label{eq:taylor_integral}
\begin{split}
    \langle \aobs \rangle_T =& \aobs(\bfzero) + 
    \sum_{\nu}\int dX_{\nu} G(X_{\nu};\sigma_{\nu,T})\sum_{i}\aobs^{i}_{\nu}X_{\nu}^{i} + \\
    & \sum_{\nu < \nu'}\int dX_{\nu} dX_{\nu'} 
       G(X_{\nu};\sigma_{\nu,T})G(X_{\nu'};\sigma_{\nu',T})
    \sum_{i,j}\aobs^{i,j}_{\nu,\nu'} X_{\nu}^{i}X_{\nu'}^{j} + ...
\end{split}
\end{equation}
The second and third terms of the right-hand side of Eq. \ref{eq:taylor_integral} are the contributions due to single-phonon processes and two-phonon processes, respectively. If multi-phonon processes are neglected we obtain,
\begin{equation}\label{eq:sp}
    \langle \aobs \rangle_T^\text{sp} = \aobs(\bfzero) + 
    \sum_{\nu}\int dX_{\nu} G(X_{\nu};\sigma_{\nu,T})\sum_{i}\aobs^{i}_{\nu}X_{\nu}^{i}
\end{equation}
We refer to this approximation as \textit{single-phonon approximation}, where each phonon mode $(\nu)$ couples independently to the electronic observable, $\aobs$, with the $i$-th order coupling constants, $\aobs^{i}_{\nu}$. The "TL-sp" approach mentioned in the main text evaluates this expression. We note that the TL method and MC sampling do not rely on the Taylor series approximation of the electronic observable ($\aobs$),  and therefore they include electron-phonon coupling up to any arbitrary order including multi-phonon effects. 

To further simplify, we invoke the quadratic approximation, i.e. we neglect terms with $i > 2$ in Eq. \ref{eq:sp}. Consequently, the phonon renormalized electronic observable becomes,
\begin{equation}\label{eq:qa_avg}
    \langle \aobs \rangle_T^{\text{sp,}\qa} = \aobs(\bfzero) +  \sum_{\nu}\frac{1}{2\omega_\nu}\aobs_{\nu}^2
    \qty[n_B(\omega_{\nu},T)+\frac{1}{2}].
\end{equation}
The frozen-phonon approach evaluates this expression, where finite differences are utilized to compute the coupling term $\aobs_{\nu}^2=\eval{\pdv[2]{\aobs}{X_{\nu}}}_0$, see Eqs. S23-S25 of reference  \citealp{Kundu_JCTC_2023}. 

\newpage

\subsection{A2. Scheme for degeneracy splitting}

\begin{figure}
    \centering
    \includegraphics[width=0.8\textwidth]{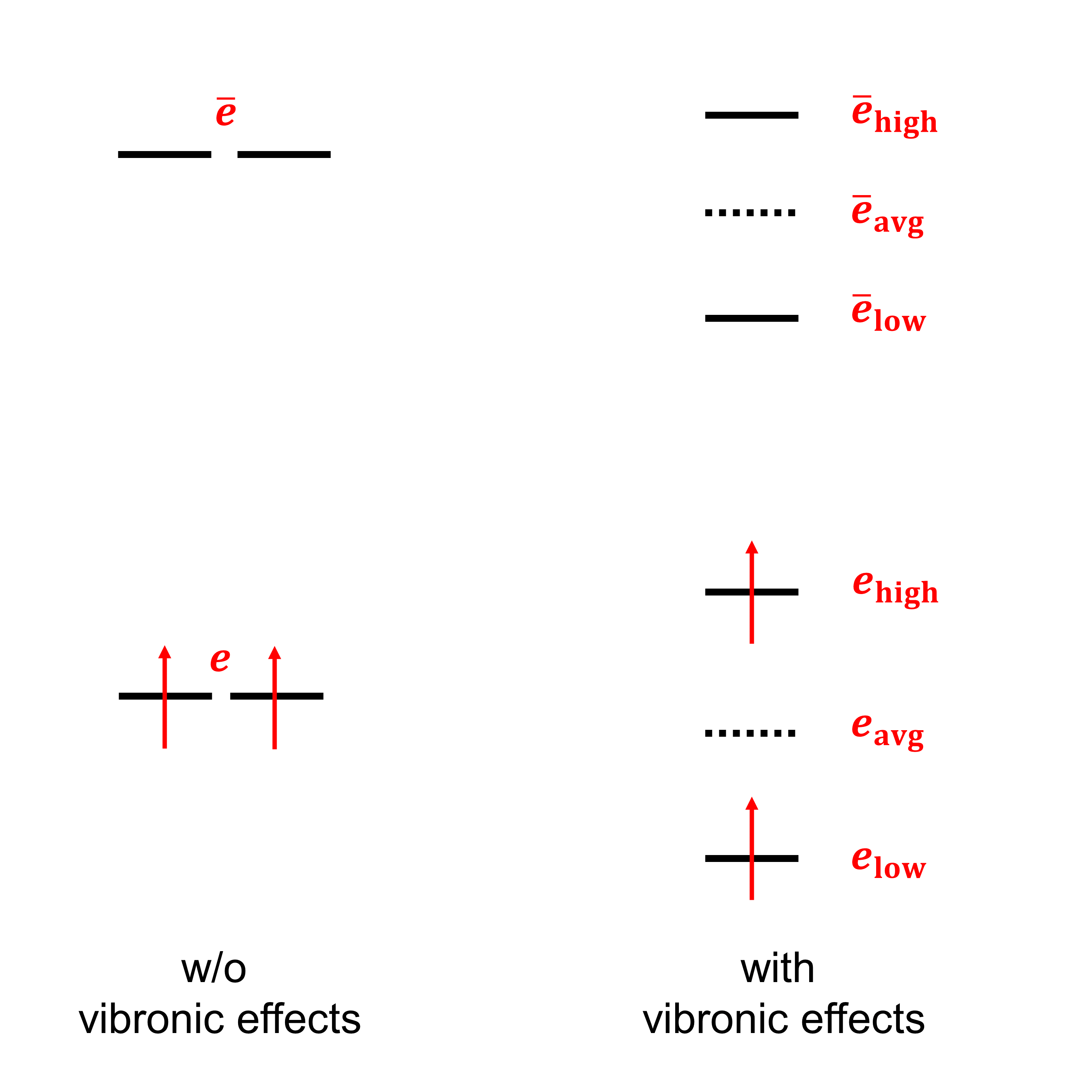}
    \caption{A scheme showing the degeneracy splitting of $e$ and $\overline{e}$ levels due to quantum vibronic effects. In the presence of such splitting, we label the $e$ and $\overline{e}$ levels with lower and higher energy as $e_\mathrm{low}$, $\overline{e}_\mathrm{low}$ and $e_\mathrm{high}$, $\overline{e}_\mathrm{high}$, respectively. After including the vibronic coupling, the barycenters (\textit{i.e} average of lower and higher energy levels) are denoted as $e_\mathrm{avg}$, $\overline{e}_\mathrm{avg}$ respectively. }
    \label{fig:orbital-splitting}
\end{figure}

\newpage

\subsection{A3. Computational Details}\label{sec:comp_details}
We carried our first principles simulations using the geometry-optimized coordinates and dynamical matrix elements obtained in previous first-principles studies.\cite{YuJin_PRM_2021,YuJin_NPJ_2022, Han_JCTC_2022} Only for the smallest supercell (C\textsubscript{62}N), we computed the dynamical matrix using the finite-difference method implemented in the PyEPFD package\cite{Kundu_PRM_2021, Kundu_JCTC_2023, pyepfd} with a symmetric Cartesian displacement of 0.005 a.u. We note that performing dynamical matrix calculations for a C\textsubscript{214}N supercell with a hybrid functional is computationally prohibitive. Therefore in an earlier study,\cite{YuJin_PRM_2021} PBE dynamical matrix elements were scaled by a factor (1.04 and 1.0514 for DDH and HSE, respectively) determined by the ratio of the highest phonon frequencies for the bulk diamond obtained with the respective hybrid and the PBE functional. This approach was found to be robust  for bulk pristine systems.\cite{YuJin_PRM_2021}.       

\begin{figure}[htbp]
    \centering

    \begin{subfigure}[b]{\textwidth}
        \centering
        \includegraphics[width=\textwidth]{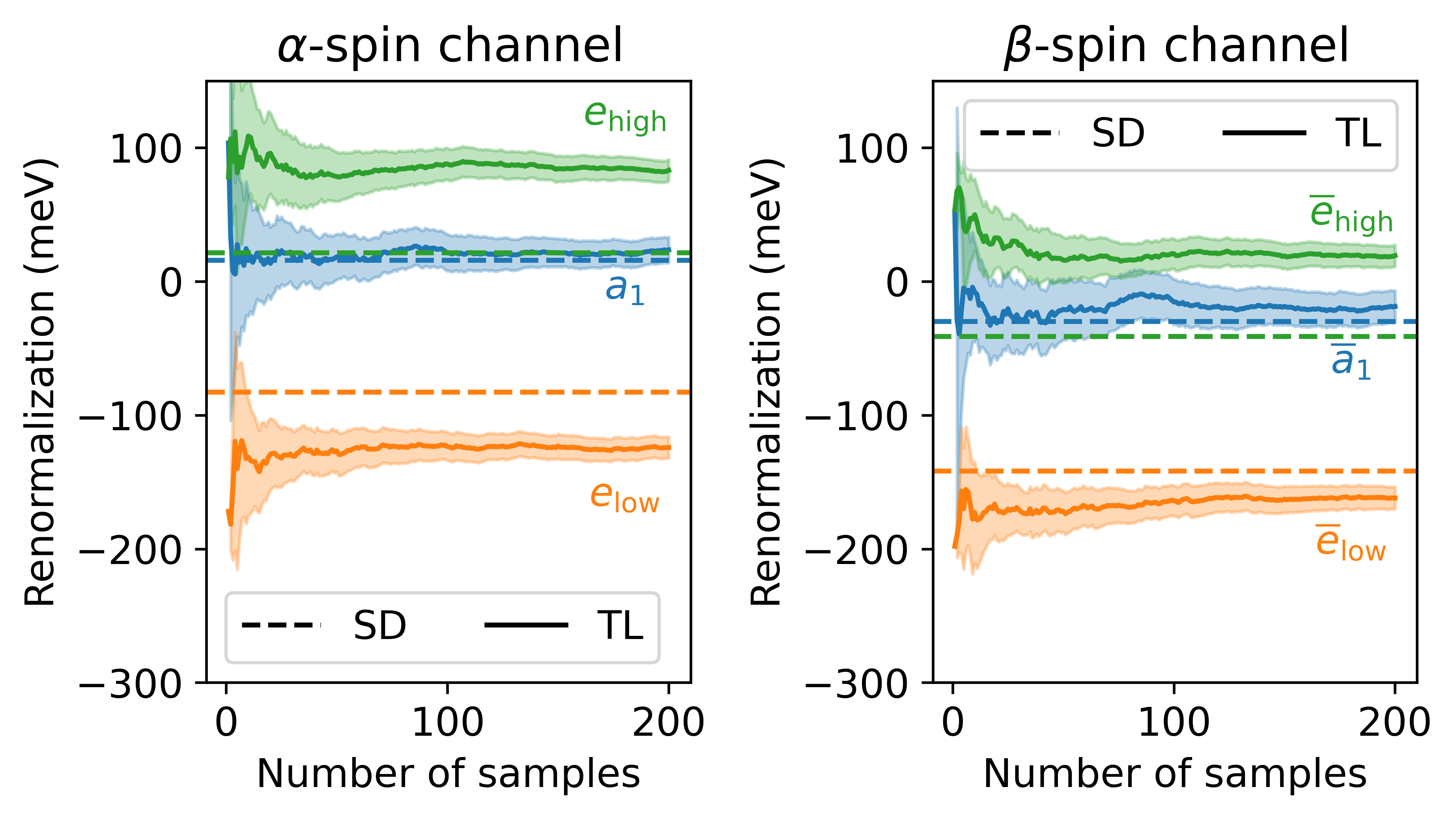}
    \end{subfigure}
    \vfill
    \begin{subfigure}[b]{\textwidth}
        \centering
        \includegraphics[width=\textwidth]{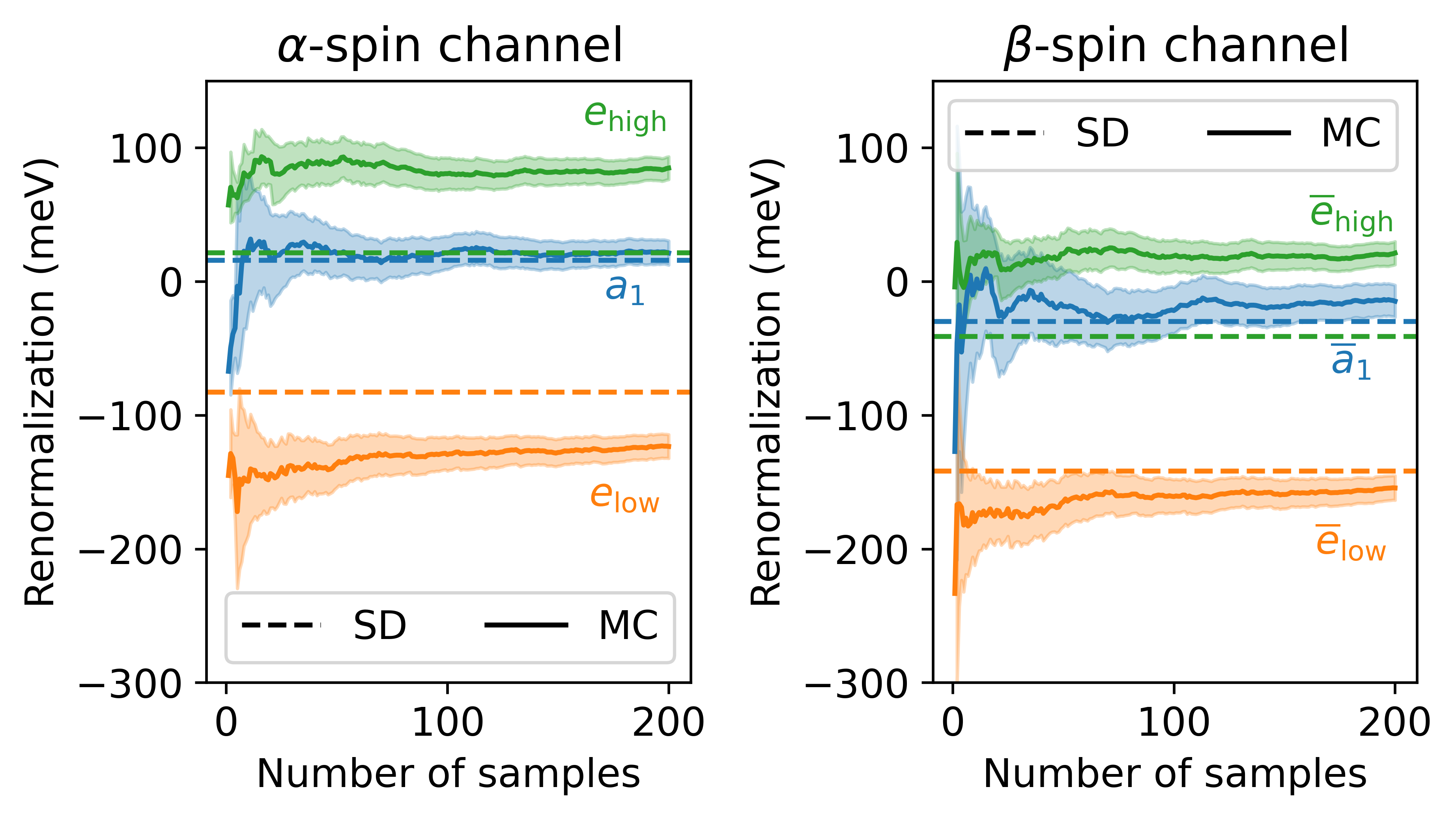}
    \end{subfigure}

    \caption{Phonon renormalizations at 300 K for various single-particle defect level energies as a function of the number of independent samples as obtained with TL sampling (top panel) and MC sampling (bottom panel) using the PBE functional on a C\textsubscript{214}N supercell. The solid lines represent the expectation values while the shaded regions signify its uncertainty (2 times the standard deviation of the mean). In each case, the renormalizations obtained with special displacement (SD) are also shown with dashed lines.}
    
    \label{fig:tl_mc_conv}
\end{figure}

Using the scaled dynamical matrix elements as input, we generated canonical ensembles of displaced configurations at a specified temperature using different stochastic algorithms: (i) special displacement (SD), \cite{Zacharias_PRB_2016} (ii) thermal line (TL) sampling,\cite{TL_Monserrat_PRB_2016} and (iii) Monte Carlo (MC) sampling.\cite{Monserrat_Rev_2018} For TL and MC samplings, using the PyEPFD\cite{Kundu_JCTC_2023, pyepfd} package, we generated 200 independent samples and their antithetic pairs producing a total of 400 configurations for each ensemble,  and we performed single-point DFT or TDDFT calculations for these configurations. Figure \ref{fig:tl_mc_conv} shows the computed phonon renormalization values at 300 K for various single-particle defect levels as a function of the number of independent samples, as obtained with TL sampling (top panel) and MC sampling (bottom panel) using the PBE functional for a C\textsubscript{214}N supercell. Note that averaging over only 100 samples (200 single-point DFT calculations) produces reasonably well-converged expectation values. For the SD method, there are only 2 configurations (when the antithetic pair is included) for each temperature and thus this technique cannot provide an estimate of stochastic uncertainties. We generated the configurations used for SD calculations with the PyEPFD\cite{Kundu_JCTC_2023, pyepfd} package,  for which we then  performed DFT single-point calculations.   

We chose one of the stochastically displaced configurations (from an ensemble with $T=300$ K) for a C\textsubscript{214}N supercell as the starting point of  first-principles molecular dynamics (FPMD) simulation with a quantum thermostat using the PBE functional and setting the thermostat temperature to 300 K. We used a time step of 0.5 fs and propagated the dynamics for 25000 steps using the i-PI---Qbox coupling scheme, \cite{Kundu_PRM_2021} where the i--PI driver\cite{ipi_Kapil_2018} moves the nuclear coordinates and the Qbox \cite{Qbox_Gygi_2008} engine provides the forces from DFT.

\begin{figure}[htb]
    \centering
    \includegraphics[width=\textwidth]{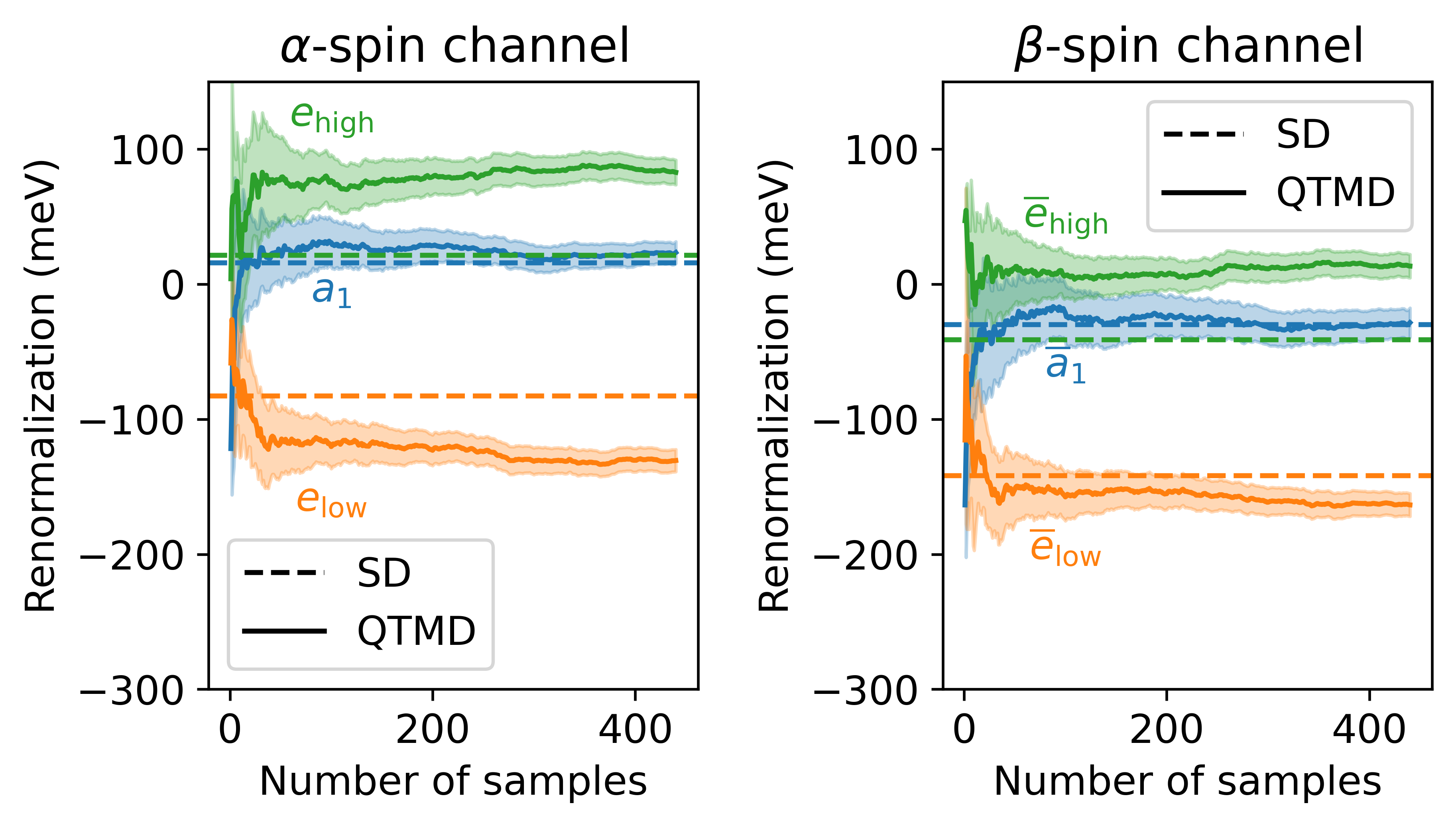}
    \caption{Phonon renormalizations at 300 K for various single-particle defect level energies as a function of the number of independent samples as obtained from a quantum thermostatted molecular dynamics (QTMD) trajectory using the PBE functional on a C\textsubscript{214}N supercell. The solid lines represent the expectation values while the shaded regions signify its uncertainty (2 times the standard deviation of the mean). In each case, the renormalizations obtained with special displacement (SD) are also shown with dashed lines.}
    \label{fig:qtmd_conv}
\end{figure}

Since we started with a configuration representative of the canonical ensemble at $T=300$ K, equilibration was attained within only 3000 steps. From the rest of the 22000  steps, we chose configurations at an interval of 50 steps (comparable to the time scale of the lowest phonon frequency) to ensure that configurations are  uncorrelated and we computed the phonon renormalization of  different defect level energies. Fig. \ref{fig:qtmd_conv} shows these renormalization values as a function of the number of independent samples drawn from the trajectory. 

\begin{figure}[htbp]
    \centering

    \begin{subfigure}[b]{\textwidth}
        \centering
        \includegraphics[width=\textwidth]{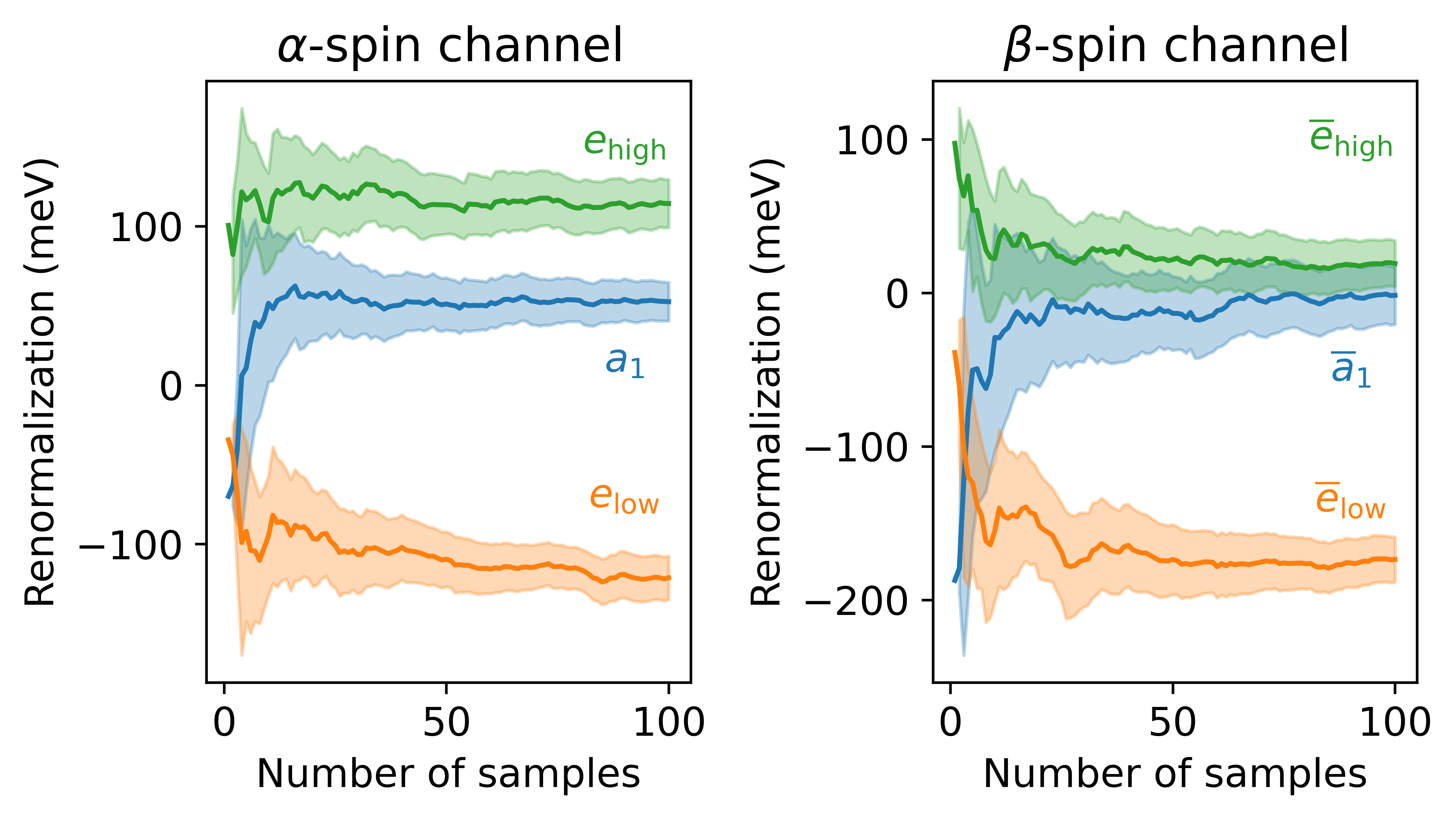}
    \end{subfigure}
    \vfill
    \begin{subfigure}[b]{\textwidth}
        \centering
        \includegraphics[width=\textwidth]{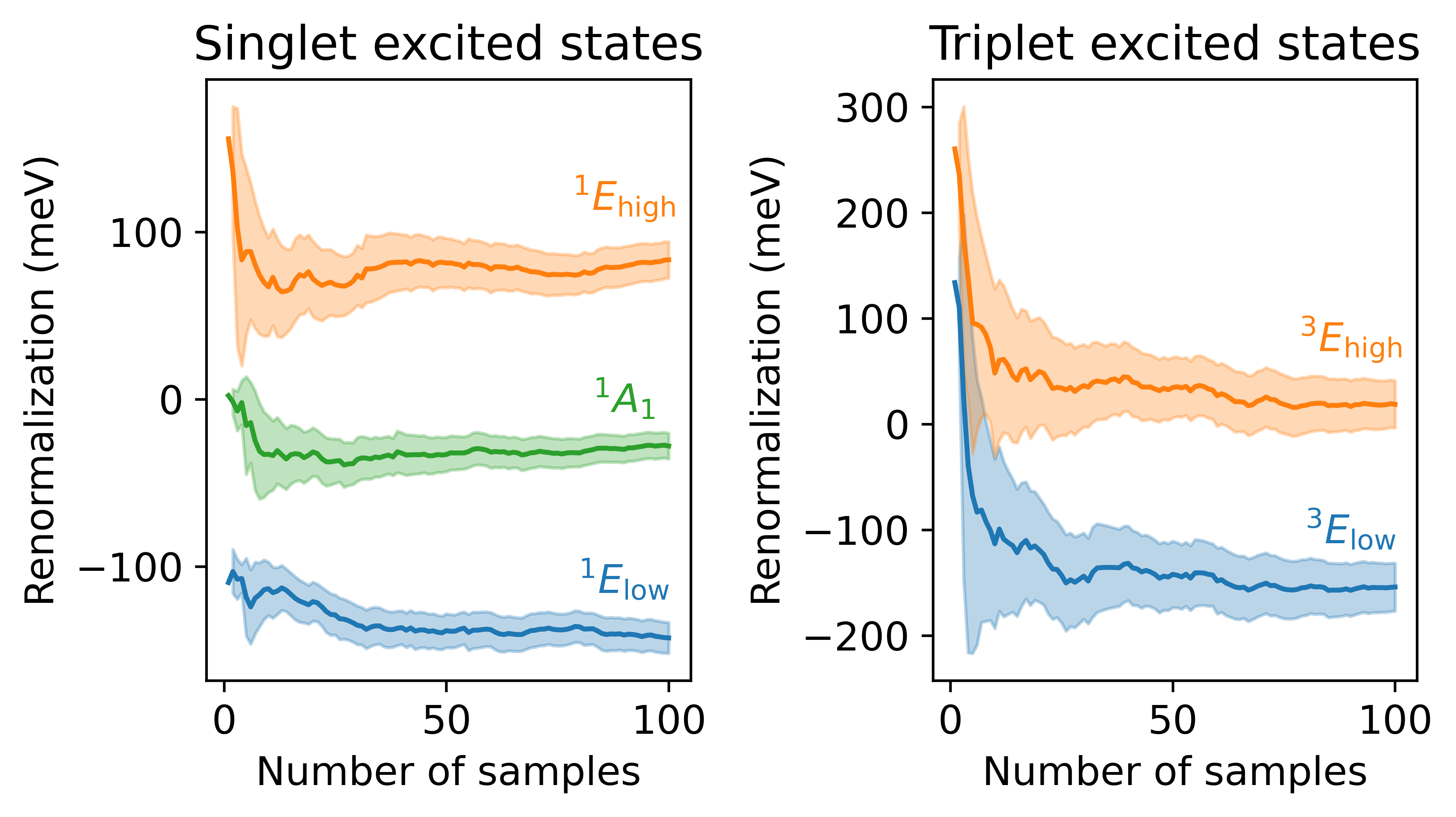}
    \end{subfigure}

    \caption{Phonon renormalizations at 300 K for various single-particle defect level energies (top panel) and VEEs (when excited from the ground state) of many-body excited states (bottom panel) as a function of the number of independent TL samples. These calculations are performed on a C\textsubscript{214}N supercell with a DDH functional with a 60Ry kinetic energy cutoff. The renormalizations of the single-particle defect levels are calculated from Kohn-Sham eigenvalues while that of the VEEs of the many-body states are obtained from TDDFT excitation. The solid lines represent the expectation values while the shaded regions signify its uncertainty (2 times the standard deviation of the mean).}
    
    \label{fig:ddh_tl_conv}
\end{figure}

Note that MD simulations with a quantum thermostat are at least a few order of magnitude costlier than stochastic simulations because at least several thousands of MD steps (DFT calculations) are required to gather a few hundred uncorrelated configurations. In contrast, for a given dynamical matrix, each configuration within a stochastic ensemble is uncorrelated; consequently, only a few hundred DFT calculations are necessary. If symmetry is used, a few hundred DFT calculations are sufficient for obtaining a dynamical matrix. In addition, the same dynamical matrix can be reused for samples at different temperatures, while separate FPMD simulations must be performed for each temperature.

For the C\textsubscript{62}N supercell, we also used frozen phonon calculations to obtain the zero-point renormalizations. Using the PyEPFD package, we computed the second derivatives of the single-particle defect level energies using a phonon mode displacement that corresponds to a potential energy change of 0.001 a.u.; see Eqs. S23-S25 of reference  \citealp{Kundu_JCTC_2023}.  

For all calculations, we used normed-conserving pseudopotentials\cite{ONCV_2015} to describe the interaction between valence and core electrons. For the C\textsubscript{214}N and C\textsubscript{510}N supercells, we used a 85 Ry kinetic energy cutoff within the Qbox settings while performing single-point DFT calculations and FPMD with a quantum thermostat. To be consistent with previous DMPT calculations, \cite{Han_JCTC_2022} we used only a 60 Ry kinetic energy cutoff for the C\textsubscript{62}N supercell. For the C\textsubscript{214}N supercell, we repeated the DFT calculations with the DDH hybrid functional and 60 Ry cutoff for  the TL sample configurations; the defect level renormalizations obtained in this way differed only by 2 meV or less from those obtained with a 85 Ry cutoff. Therefore for all TDDFT calculations, we used a 60 Ry kinetic energy cutoff. We used the TDDFT implementation \cite{YuJin_TDDFT_2023} of the WEST package\cite{west1, west2} after obtaining the ground state wavefunctions with the Quantum Espresso package\cite{qe1, qe2}.

Although for the TL sampling of  a C\textsubscript{214}N supercell with a hybrid functional (DFT/TDDFT) we could afford only 100 independent samples and their antithetic pairs (200 single-point calculations), we found that the convergence  of  the various renormalization values was sufficiently good, as shown  in Fig. \ref{fig:ddh_tl_conv}.

\newpage

\subsection{A4. Notes on Special Displacement Method}

\begin{table}[H]
    \centering
    \caption{Comparison of band gap renormalizations (meV) at 300 K computed using different supercells for the NV$^-$ center in diamond, and the PBE functional, as obtained with different sampling methods.}
    \begin{tabular}{ccccc}
    \toprule
    Supercell & SD & TL & MC & MD \\
    \midrule
     C\textsubscript{214}N & $-311$ & $-319$ & $-319$ & $-322$  \\
     C\textsubscript{510}N  & $-333$ & $-336$ & $-336$ &   \\
     \bottomrule
    \end{tabular}
    \label{tab:band_gap}
\end{table}

\begin{figure}[htb]
\centering
\includegraphics[width=9.5cm]{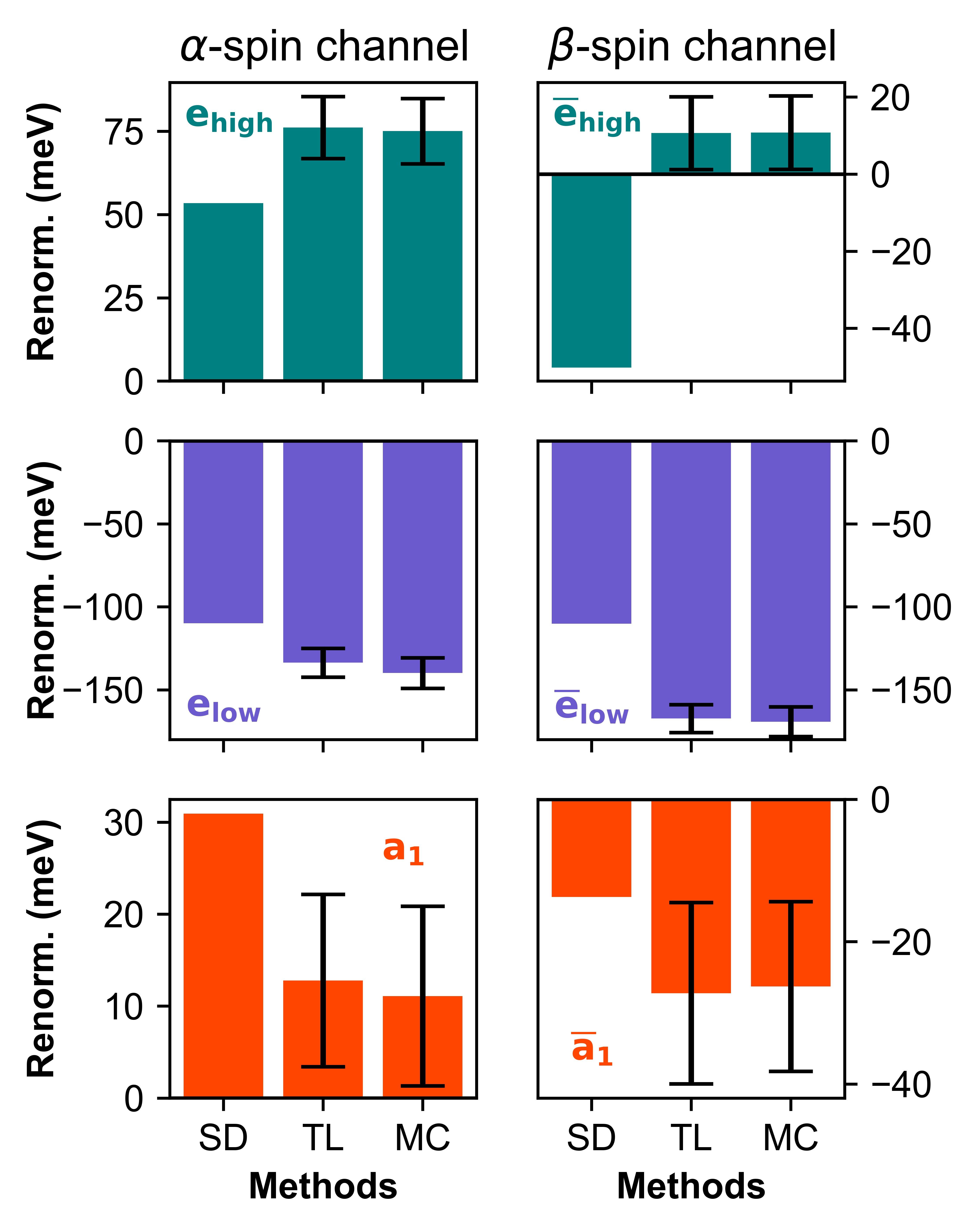}
\caption{The ground state phonon renormalizations of the single-particle defect levels ($a_1$, $\overline{a}_1$, $e_{\mathrm{low}}$, $e_{\mathrm{high}}$, $\overline{e}_{\mathrm{low}}$, $\overline{e}_{\mathrm{high}}$) at 300 K obtained with various sampling techniques: special displacement (SD), thermal line (TL), and Monte Carlo (MC). For these calculations, we used the PBE functional and a C\textsubscript{510}N supercell. Black bars indicate the error bars.}
\label{fig:methods_compare_nv512}
\end{figure}
In the limit of large supercells, the phonon renormalizations obtained with the special displacement (SD) method are expected to converge to those obtained with thermal line sampling (TL) or Monte Carlo (MC) methods.\cite{Zacharias_PRB_2016, Zacharias_PRR_2020} 
In the case of the delocalized band edges, the band gap obtained with the SD method is within 2.5\% of the TL(MC) results even for a C\textsubscript{214}N supercell and the agreement improves for a larger (C\textsubscript{510}N) supercell, see Table \ref{tab:band_gap}.
However, for the renormalizations for the localized defect levels, even for a C\textsubscript{510}N supercell, the SD results are not yet converged to those of the TL(MC) method, see Fig. \ref{fig:methods_compare_nv512}

\newpage
\subsection{A5. Finite Size Scaling Analysis}\label{sec:fss}

\begin{figure}[htb]
    \centering
    \begin{subfigure}[b]{0.54\textwidth}
        \centering
        \includegraphics[width=\textwidth]{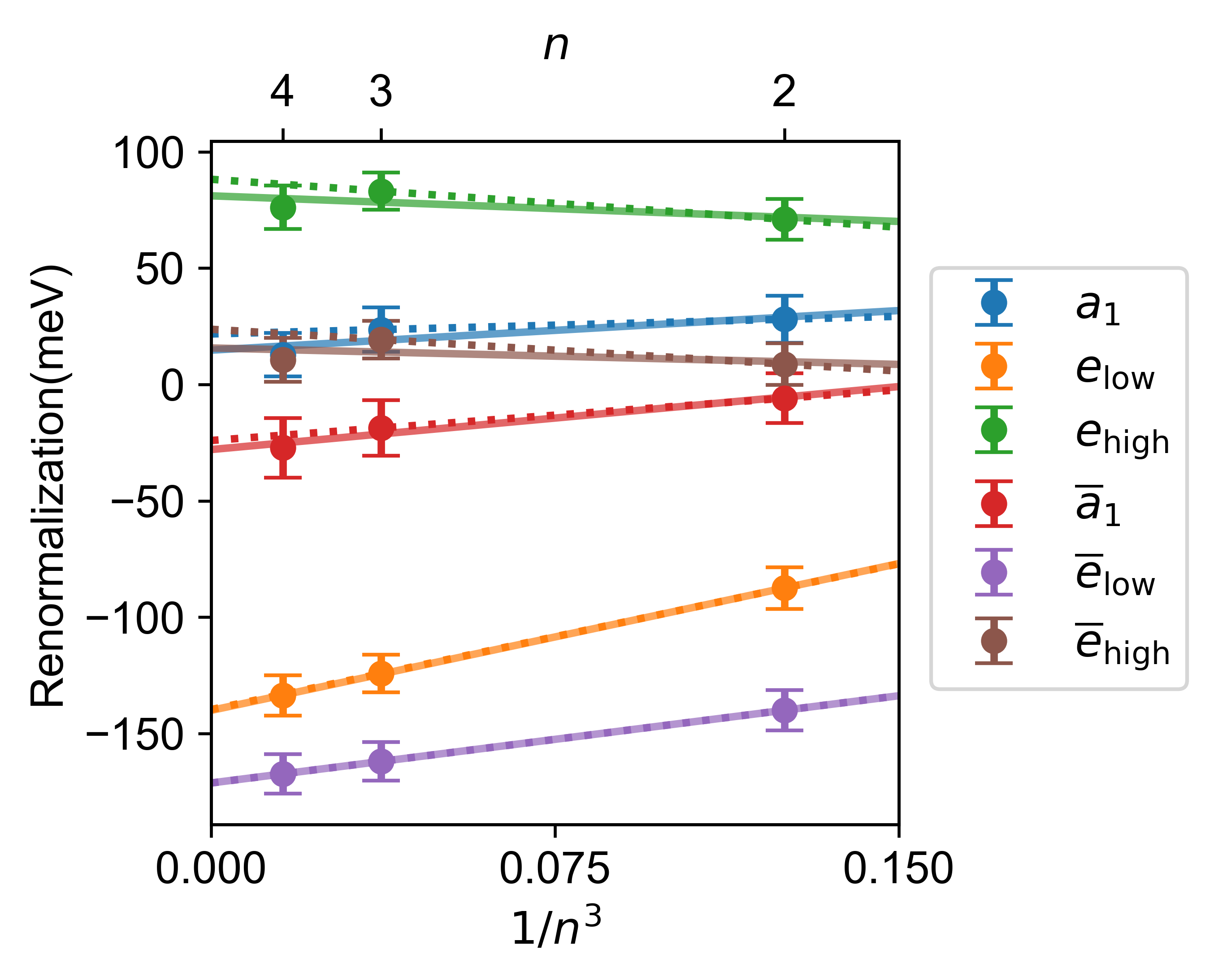}
    \end{subfigure}
    \hspace{0.1cm}
    \begin{subfigure}[b]{0.43\textwidth}
        \centering
        \includegraphics[width=\textwidth]{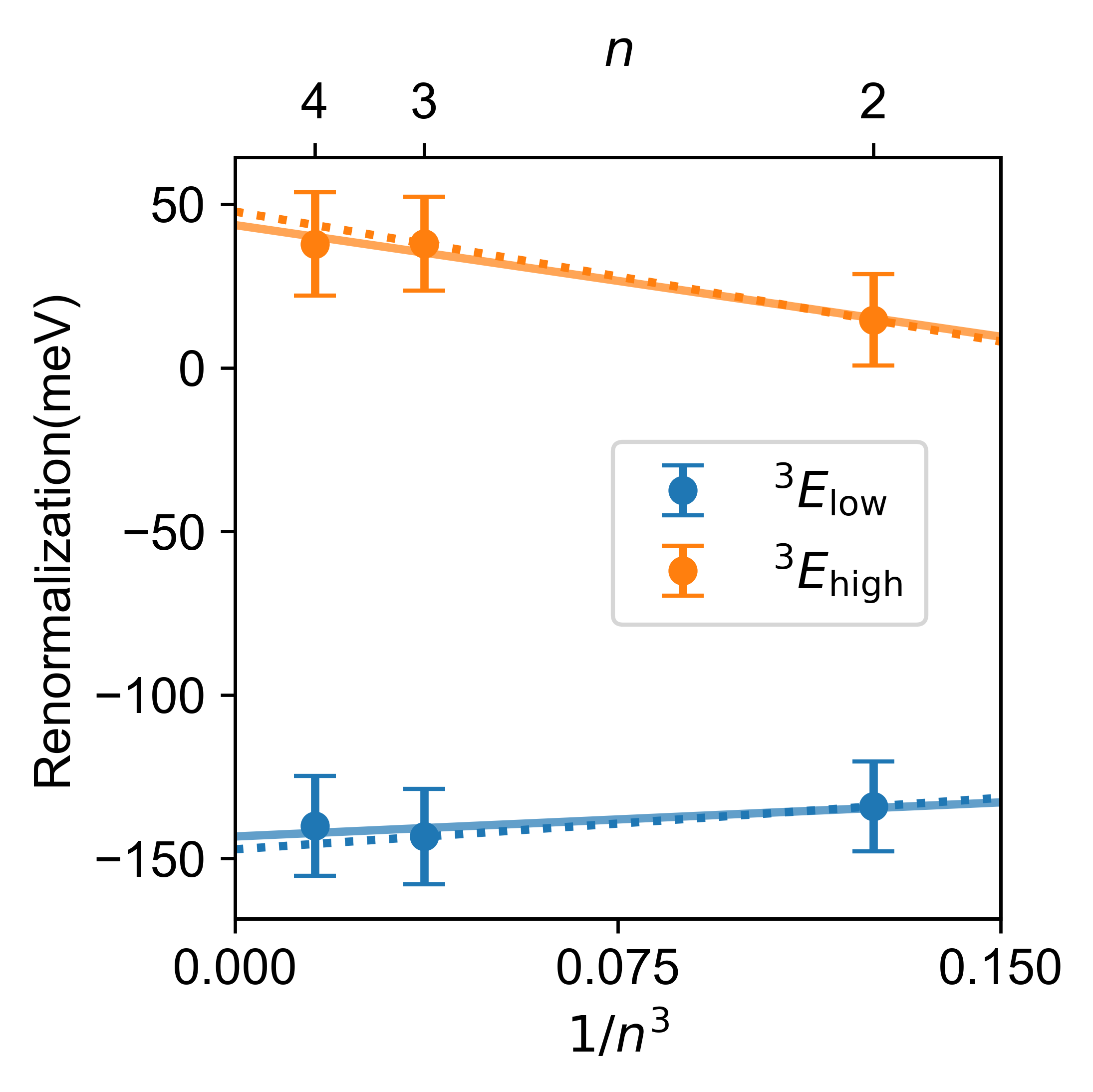}
    \end{subfigure}
    \caption{ Phonon renormalizations (PBE) of the single-particle defect level energies (left panel) and the VEEs of the $^3E$ state (right panel) for various $n \cross n \cross n$ supercells. VEE renormalization for the $^3E_\mathrm{low}$($^3E_\mathrm{high}$) state is derived from the difference in renormalizations of the $\overline{e}_\mathrm{low}$($\overline{e}_\mathrm{high}$) and $\overline{a}_1$ levels. These results are linearly extrapolated, see Eq. \ref{eq:extrapol}, to infinite supercell limit using the results obtained only with $n=2,3$ (dotted lines) and with $n=2,3,4$ (solid lines).}
    \label{fig:fss_pbe}
\end{figure}

Table \ref{tab:fss_eigval_renorm} shows the phonon renormalization values of the single-particle defect levels obtained using TL sampling at 300 K for $n \cross n \cross n$ supercells with $n=2,3,4$. These calculations can be extrapolated to the thermodynamic limit $(n \rightarrow \infty)$ using the following equation:
\begin{equation}
    \Delta E_n = \Delta E_{\infty}+B/n^3
    \label{eq:extrapol}
\end{equation}
where $\Delta E_n$ denotes the phonon renormalization obtained with an $n \cross n \cross n$ supercell. We performed the linear extrapolation in two ways: (i) using the data points only for $n=2,3$ and (ii) using all data points (i.e., $n=2, 3, 4$). Fig. \ref{fig:fss_pbe} shows that the resulting linear fits are virtually the same and the difference between the  extrapolated values is smaller than the stochastic uncertainties for each data point. This indicates that the extrapolation using the results obtained only with $n=2,3$ is meaningful. We used such 2-point linear extrapolations for the results obtained with hybrid functionals.  

\begin{table}[H]
    \centering
    \caption{ Phonon renormalizations (meV) at 300 K for the single particle defect level energies obtained with TL sampling for the different $n \cross n\cross n$ supercells as well as the extrapolated value in the thermodynamic ($n \rightarrow \infty$) limit}.
    \begin{threeparttable}
    \begin{tabular}{cccccc}
    \toprule
        Functional & Level &  C\textsubscript{62}N $(n=2) \tnote{a}$  & C\textsubscript{214}N $(n=3)$ \tnote{b} & C\textsubscript{510}N $(n=4)$ \tnote{c} &  $n \rightarrow \infty$ \tnote{d} \\
    \midrule      
        PBE & $a_1$ & 28 ± 10 & 24 ± 10 & 13 ± 9 & 15 \\
            & $e_\mathrm{low}$ & -87 ± 9 & -124 ± 8  & -134 ± 9 & -140 \\
            & $e_\mathrm{high}$ & 71 ± 9 & 83 ± 8 &  76 ± 9 & 81 \\
            & $\overline{a}_1$ & -6 ± 11 & -19 ± 12 & -27 ± 13 & -28 \\
            & $\overline{e}_\mathrm{low}$ & -140 ± 9 & -162 ± 8 & -167 ± 8 & -171 \\
            & $\overline{e}_\mathrm{hight}$ & 9 ± 9 & 19 ± 8 & 11 ± 9 & 16 \\
    \midrule
        DDH & $a_1$ & 55 ± 11 & 52 ± 12 & --- & 51 \\
            & $e_\mathrm{low}$ & -90 ± 9 & -121 ± 14  & --- & -135 \\
            & $e_\mathrm{high}$ & 108 ± 10 & 114 ± 15 &  --- & 117  \\
            & $\overline{a}_1$ & -6 ± 12 & -2 ± 19 & --- & 0 \\
            & $\overline{e}_\mathrm{low}$ & -151 ± 10 & -174 ± 15  & --- & -183 \\
            & $\overline{e}_\mathrm{hight}$ & 22 ± 10 & 19 ± 15 & --- & 18 \\
    \midrule
        HSE & $a_1$ & --- & 60 ± 13 & --- & --- \\
            & $e_\mathrm{low}$ & --- & -137 ± 16  & --- & --- \\
            & $e_\mathrm{high}$ & --- & 108 ± 14 &  --- & ---  \\
            & $\overline{a}_1$ & --- & -24 ± 23 & --- & --- \\
            & $\overline{e}_\mathrm{low}$ & --- & -187 ± 13  & --- & --- \\
            & $\overline{e}_\mathrm{hight}$ & --- & 20 ± 14 & --- & --- \\
    \bottomrule
    \end{tabular}
    \begin{tablenotes}
        \item[a] DFT single-point calculations were performed on 400 TL samples using a 60 Ry kinetic energy cutoff.
        \item[b] DFT/PBE single-point calculations were performed on 400 TL samples, while DFT/hybrid calculations were performed on 200 TL samples. 85 Ry kinetic energy cutoff is used for PBE and HSE while 60 Ry cutoff is used for DDH.
        \item[c] DFT/PBE single-point calculations were performed on 400 TL samples using a 85 Ry kinetic energy cutoff.
        \item[d] Extrapolated using Eq. \ref{eq:extrapol}. In Fig-4A of the main text, uncertainties obtained with the largest supercell are reported.
    \end{tablenotes}
    \end{threeparttable}
    \label{tab:fss_eigval_renorm}
\end{table}

\begin{table}[H]
    \centering
    \caption{TDDFT/DDH VEEs, its phonon renormalizations ($\Delta$VEE) at 300 K, and the renormalized VEEs (VEE + $\Delta$VEE) for excited states (when excited from the ground state) as obtained with different $n \cross n\cross n$ supercells. All energies are in eV.}
    \begin{threeparttable}
    \begin{tabular}{ccccc}
    \toprule
       State  & Quantity  & C\textsubscript{62}N $(n=2)$\tnote{a}  & C\textsubscript{214}N $(n=3)$\tnote{b} & $n \rightarrow \infty$\tnote{c} \\
    \midrule 
        $^1E_\mathrm{low}$  & VEE                &  0.655         &  0.678         &  0.679 \\
                            & $\Delta$VEE        & -0.124 ± 0.005 & -0.143 ± 0.009 & -0.150 \\
                            & VEE + $\Delta$VEE  &  0.531 ± 0.005 &  0.535 ± 0.009 &  0.529 \\
    \midrule
        $^1E_\mathrm{high}$ & VEE                &  0.655         &  0.678         &  0.679 \\
                            & $\Delta$VEE        &  0.078 ± 0.007 &  0.083 ± 0.011 &  0.085 \\
                            & VEE + $\Delta$VEE  &  0.733 ± 0.007 &  0.761 ± 0.011 &  0.764 \\
    \midrule
        $^1A_1$             & VEE                &  1.866         &  1.955         &  1.960 \\
                            & $\Delta$VEE        & -0.044 ± 0.007 & -0.028 ± 0.008 & -0.021 \\
                            & VEE + $\Delta$VEE  &  1.822 ± 0.007 &  1.927 ± 0.008 &  1.939 \\
    \midrule
        $^3E_\mathrm{low}$  & VEE                &  2.251         &  2.391         &  2.398 \\
                            & $\Delta$VEE        & -0.139 ± 0.014 & -0.154 ± 0.023 & -0.160 \\
                            & VEE + $\Delta$VEE  &  2.112 ± 0.014 &  2.237 ± 0.023 &  2.238 \\
    \midrule
        $^3E_\mathrm{low}$  & VEE                &  2.251         &  2.391         &  2.398 \\
                            & $\Delta$VEE        &  0.017 ± 0.014 &  0.019 ± 0.022 &  0.020 \\
                            & VEE + $\Delta$VEE  &  2.268 ± 0.014 &  2.410 ± 0.022 &  2.418 \\
    \bottomrule
    \end{tabular}
    \begin{tablenotes}
        \item[a] 400 configurations were used to determine phonon renormalizations.
        \item[b] 200 configurations were used to determine phonon renormalizations.
        \item[c] Eq. \ref{eq:extrapol} is used for the extrapolation. In Table 3 of the main text, uncertainties obtained with the largest supercell are reported. 
    \end{tablenotes}
    \end{threeparttable} 
    \label{tab:tddft_all}
\end{table}

\newpage
\section{A6. {\it T}-dependence of phonon renormalizations: Effect of functionals and supercells}\label{sec:t_dependence}

\begin{figure}[htbp]
    
    \includegraphics[width=\textwidth]{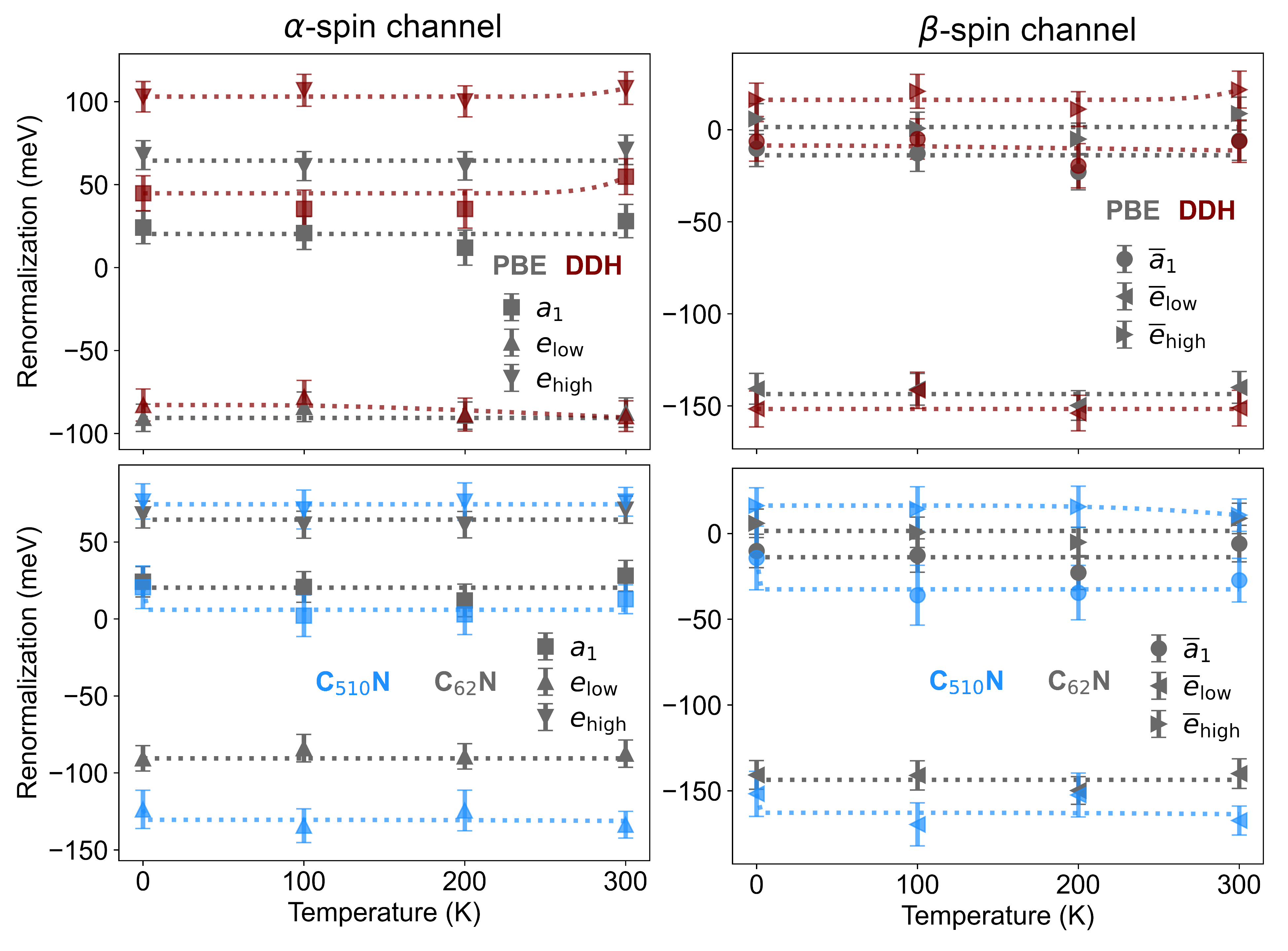}
    \caption{Temperature dependence of the defect level renormalizations. Top panel compares PBE and DDH functionals for a C\textsubscript{62}N supercell. Bottom panel compares C\textsubscript{62}N and C\textsubscript{510}N supercells using the PBE functional. For each temperature, 400 configurations obtained from a TL sampling were used. Symbols and lines represent the TL sampling results and their Vi\~{n}a model\cite{Vina_PRB_1984} fit. }
    \label{fig:defect}
\end{figure}

Figure \ref{fig:defect}, top panel, compares the defect-level renormalizations obtained with the PBE and DDH functionals. Despite PBE and DDH predicting different values for a specific defect level, both agree on the negligible temperature dependence within 0---300 K. Figure \ref{fig:defect}, bottom panel, compares size effects on these defect-level renormalizations; in spite of size effects being present, none of our calculations predict any temperature dependence within 0---300 K.

\begin{figure}[htbp]
    \centering

    \begin{subfigure}[b]{0.6\textwidth}
        \centering
        \includegraphics[width=\textwidth]{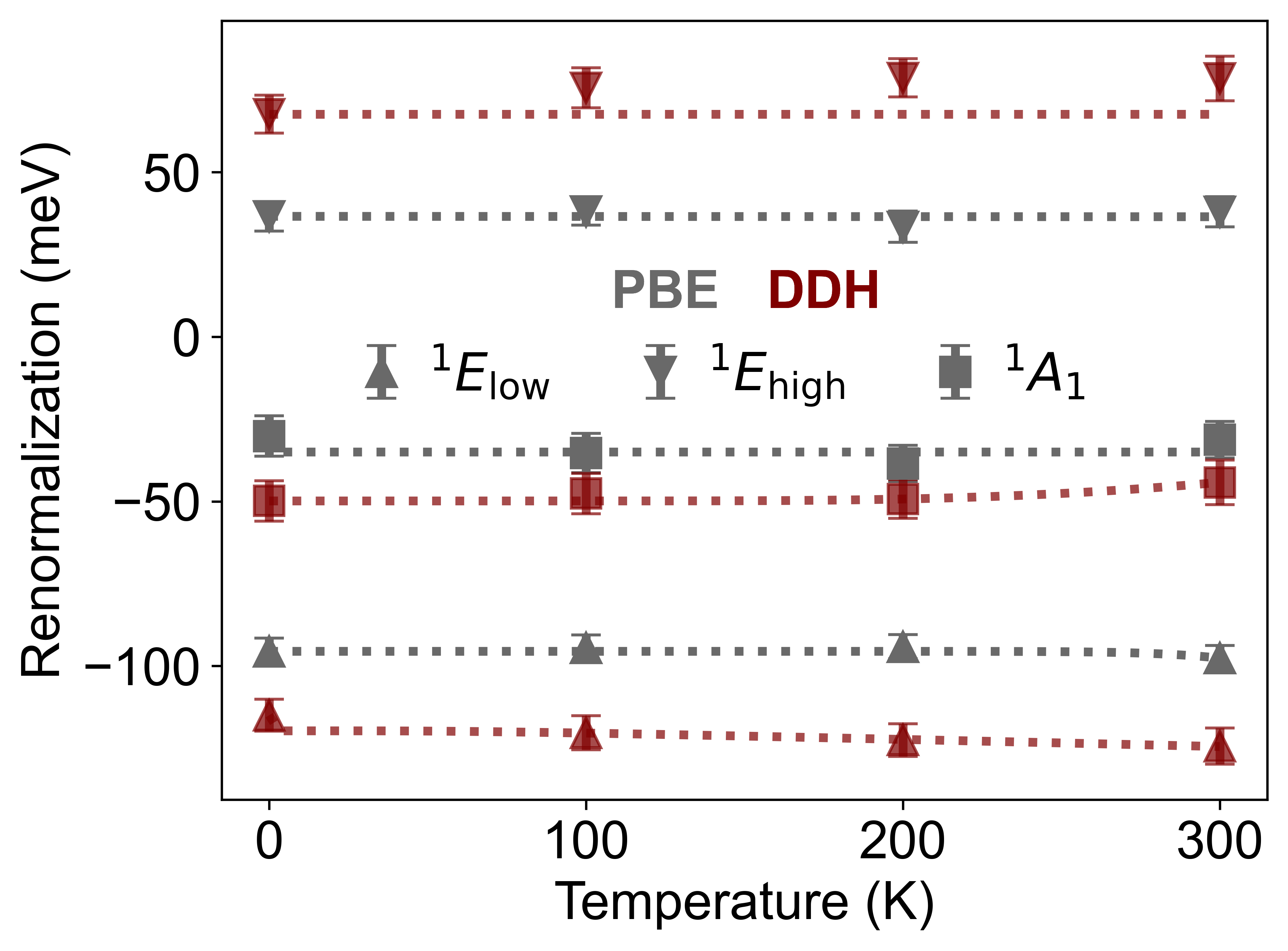}
    \end{subfigure}
    \vspace{0.05cm}
    \begin{subfigure}[b]{0.6\textwidth}
        \centering
        \includegraphics[width=\textwidth]{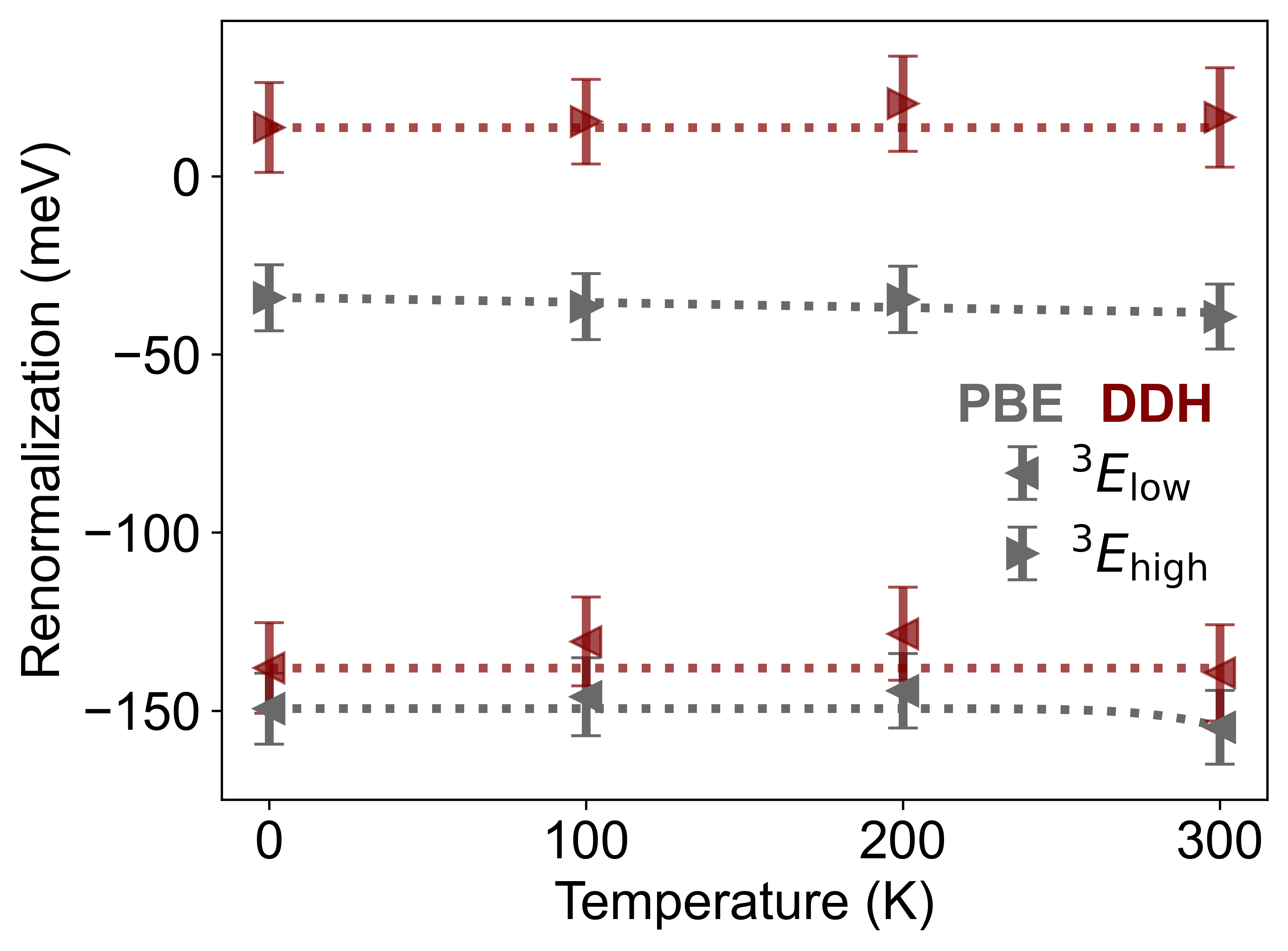}
    \end{subfigure}
    \caption{Comparison of the PBE and DDH functionals highlighting the $T$-dependence of VEE renormalizations of singlet (top panel) and triplet (bottom panel) excited states when excited from the ground state. For each temperature, 400 displaced configurations for a C\textsubscript{62}N supercell were generated using the TL sampling and single-point TDDFT calculations followed. Symbols and lines represent the TL sampling results and their Vi\~{n}a model\cite{Vina_PRB_1984} fit.}
    \label{fig:excited}
\end{figure}

Figure \ref{fig:excited} compares PBE and DDH functional results highlighting the $T$-dependence of VEE renormalizations of singlet (top panel) and triplet (bottom panel) excitations from the ground state. Despite PBE and DDH functionals predicting different values for the renormalizations of  specific excited states, both functional yield  a negligible temperature dependence within 0---300 K.

\clearpage

\begin{acknowledgement}
We thank Yu Jin for providing the force constants for C\textsubscript{214}N and C\textsubscript{510}N supercells and for many useful discussions.
This work was supported by MICCoM, as part of the Computational Materials Sciences Program funded by the U.S. Department of Energy, Office of Science, Basic Energy Sciences, Materials Sciences, and Engineering
Division through Argonne National Laboratory, under Contract No. DE-AC02-06CH11357. This research used the resources of the University of Chicago Research Computing Center.
\end{acknowledgement}

\bibliography{references}

\end{document}